\documentclass[twocolumn,showpacs,amsmath,amssymb,prd,floatfix]{revtex4}

\usepackage{graphicx}
\usepackage{dcolumn}
\usepackage[varg]{txfonts}
\usepackage{bm}

\begin{document}

\title{Integrated Perturbation Theory and One-loop Power Spectra
  of Biased Tracers}

\author{Takahiko Matsubara} \email{taka@kmi.nagoya-u.ac.jp}
\affiliation{%
  Department of Physics, Nagoya University, Chikusa, Nagoya, 464-8602,
  Japan;}%
\affiliation{%
  Kobayashi-Maskawa Institute for the Origin of Particles and the
  Universe, Nagoya University, Chikusa, Nagoya, 464-8602, Japan}%

\date{\today}

\begin{abstract}
  General and explicit predictions from the integrated perturbation
  theory (iPT) for power spectra and correlation functions of biased
  tracers are derived and presented in the one-loop approximation. The
  iPT is a general framework of the nonlinear perturbation theory of
  cosmological density fields in presence of nonlocal bias,
  redshift-space distortions, and primordial non-Gaussianity. Analytic
  formulas of auto and cross power spectra of nonlocally biased
  tracers in both real and redshift spaces are derived and the results
  are comprehensively summarized. The main difference from previous
  formulas derived by the present author is to include effects of
  generally nonlocal Lagrangian bias and primordial non-Gaussianity,
  and the derivation method of the new formula is fundamentally
  different from the previous one. Relations to recent work on
  improved methods of nonlinear perturbation theory in literature are
  clarified and discussed.
\end{abstract}

\maketitle


\section{\label{sec:intro}
Introduction
}

Density fluctuations in the universe contain invaluable information on
cosmology. For example, the history and ingredients of the universe
are encoded in detailed patterns of the density fluctuations. The
large-scale structure (LSS) of the universe is one of the most popular
ways to probe the density fluctuations in the universe. Spatial
distributions of galaxies and other astronomical objects which can be
observed reflect the underlying density fluctuations in the universe.

In cosmology, it is crucial to investigate the spatial distributions
of dark matter, which dominates the mass of the universe.
Unfortunately, distributions of dark matter are difficult to directly
observe, because the only interaction we know that the dark matter
surely has is the gravitational interaction. Consequently, we need to
estimate the density fluctuations of the universe by means of indirect
probes such as galaxies, which have electromagnetic interactions.

Relations between distributions of observable objects and those of
dark matter are nontrivial. On very large scales where the linear
theory can be applied, the relations are reasonably represented by the
linear bias; the density contrasts of dark matter $\delta_{\rm m}$ and
those of observable objects $\delta_X$ are proportional to each other,
$\delta_X = b \delta_{\rm m}$, where $b$ is a constant called the bias
parameter. However, nonlinear effects cannot be neglected when we
extract cosmological information as much as possible from
observational data of LSS, and bias relations in nonlinear regime are
not as simple as those in linear regime.

Observations of LSS play an important role in cosmology. Shapes of
power spectra of galaxies and clusters contain information on the
density parameters of cold dark matter $\varOmega_{\rm CDM}$, baryons
$\varOmega_{\rm b}$ and neutrinos $\varOmega_\nu$ in the universe.
Precision measurements of baryon acoustic oscillations (BAO) in galaxy
power spectra or correlation functions can constrain the nature of
dark energy \cite{EHT98,mat04,eis05}, which is a driving force of the
accelerated expansion of the present universe. The non-Gaussianity in
the primordial density field induces a scale-dependent bias in biased
tracers of LSS on very large scales
\cite{dal08,MV08,slo08,TKM08,DS09}. Cosmological information contained
in detailed features in LSS is so rich that there are many ongoing and
future surveys of LSS, such as BOSS \cite{daw13}, FMOS FastSound
\cite{FastSound}, BigBOSS \cite{sch11}, LSST \cite{LSST09}, Subaru PFS
\cite{PFS12}, DES \cite{DES}, Euclid \cite{Euclid11}, etc.

Elucidating nonlinear effects on observables in LSS has crucial
importance in the precision cosmology. While strongly nonlinear
phenomena are difficult to analytically quantify, the perturbation
theory is useful in understanding quasi-nonlinear regime. The
traditional perturbation theory describes evolutions of mass density
field on large scales where the density fluctuations are small.
However, spatial distributions of astronomical objects such as
galaxies do not exactly follow the mass density field, and they are
biased tracers. Formation processes of astronomical objects are
governed by strongly nonlinear dynamics including baryon physics etc.,
which cannot be straightforwardly treated by the traditional
perturbation theory.

Even though the tracers are produced through strongly nonlinear
processes, it is still sensible to apply the perturbation theory to
study LSS on large scales. For example, the biasing effect in linear
theory is simply represented by a bias parameter $b$ as described
above. However, biasing effects in higher-order perturbation theory
are not that simple. A popular model of the biasing in the context of
nonlinear perturbation theory is the Eulerian local bias
\cite{HMV98,SCF99,tar00,mcd06,JK08}. This model employs freely fitting
parameters in every orders of perturbations, and is just a
phenomenological model because the Eulerian bias is not definitely
local in reality.

The integrated perturbation theory (iPT) \cite{mat11} is a framework
of the perturbation theory to predict observable power spectra and any
higher-order polyspectra (or the correlation functions) of nonlocally
biased tracers. In addition, the effects of redshift-space distortions
and primordial non-Gaussianity are naturally incorporated. This theory
is general enough so that any model of nonlocal bias can be taken into
account. Precise mechanisms of bias are still not theoretically
understood well, and are under active investigations. The framework of
iPT separates the known physics of gravitational effects on spatial
clustering from the unknown physics of complicated bias. The unknown
physics of nonlocal bias is packed into ``renormalized bias
functions'' $c_X^{(n)}$ in the iPT formalism. Once the renormalized
bias functions are modeled for observable tracers, weakly nonlinear
effects of gravitational evolutions are taken care of by the iPT. The
iPT is a generalization of a previous formulation called Lagrangian
resummation theory (LRT) \cite{mat08a,mat08b,oka11,SM11} in which
only local models of Lagrangian bias can be incorporated.

In recent developments, the model of bias from the halo approach has
turned out to be quite useful in understanding the cosmological
structure formations \cite{PS74,BCEK91,MW96,MJW97,ST99,SSHJ01,CS02}.
The halo bias is naturally incorporated in the framework of iPT.
Predictions of iPT combined with the halo model of bias do not contain
any fitting parameter once the mass function and physical mass of
halos are specified. This property is quite different from other
phenomenological approaches to combine the perturbation theory and
bias models.

A concept of nonlocal Lagrangian bias has recently attracted
considerable attention \cite{CSS12,CS12,SCS12}. Extending the halo
approach, a simple nonlocal model of Lagrangian bias is recently
proposed \cite{mat12} for applications to the iPT. Applying this
nonlocal model of halo bias to evaluating the scale-dependent bias in
the presence of primordial non-Gaussianity, not only the results of
peak-background split are reproduced, but also more general formula is
obtained. In this paper, the usage of this simple model of nonlocal
halo bias in the framework of iPT is explicitly explained.

The bias in the framework of iPT does not have to be a halo bias.
There are many kinds of tracers for LSS, such as various types of
galaxies, quasars, Ly-$\alpha$ absorption lines, 21cm absorption and
emission lines, etc. Once the bias model for each kind of objects is
given, it is straightforward to calculate biased power spectra and
polyspectra of those tracers in the framework of iPT. As described
above, it is needless to say that detailed mechanisms of bias for
those tracers have not been fully understood yet. As emphasized above,
the iPT separates the difficult problems of fully nonlinear biasing
from gravitational evolutions in weakly nonlinear regime.

While the basic formulation of iPT is developed in Ref.~\cite{mat11},
explicit calculations of the nonlinear power spectra are not given in
that reference. The purpose of this paper is to give explicit
expressions of biased power spectra with an arbitrary model of
nonlocal bias in the one-loop approximations, in which leading-order
corrections to the nonlinear evolutions are included. The expressions
are given both in real space and in redshift space. Three-dimensional
integrals in the formal expressions of one-loop power spectra are
reduced to one- and two-dimensional integrals, which are easy and
convenient for numerical integrations. Contributions from primordial
non-Gaussianity are also taken into account in the general
expressions. Explicit formulas of the renormalized bias functions are
provided for a simple model of nonlocal halo bias. In this way,
general formulas of power spectra of biased objects in the one-loop
approximation is provided in this paper.

Since the iPT framework is based on the Lagrangian perturbation theory
(LPT) \cite{buc89,mou91,buc92,cat95,RB12,tat13}, a scheme of
resummations of higher-order perturbations in terms of the Eulerian
perturbation theory (EPT) \cite{BCGS02} is naturally considered
\cite{mat08a}. In this paper, we clarify the relations of the present
formula of iPT and some previous methods of resummation technique such
as the renormalized perturbation theory \cite{CS06a,CS06b}, the Gamma
expansions \cite{BCS08,BCS10,BCS12,CSB12,TBNC12,TNB13}, the Lagrangian
resummation theory \cite{mat08a,mat08b,oka11,SM11}, and the
convolution perturbation theory \cite{CRW13}. Some aspects for the
future developments of iPT are suggested.

This paper is organized as follows. In Sec.~\ref{sec:PSiPT}, formal
expressions of power spectra in the framework of iPT with an arbitrary
model of bias are derived. A simple model of renormalized bias
functions for a nonlocal Lagrangian bias in the halo approach are
summarized. In Sec.~\ref{sec:ExplicitFormulas}, explicit formulas of
biased power spectra, which are the main results of this paper, are
derived and presented. Relations to other previous work in literature
are clarified in Sec.~\ref{sec:RelationPrevious}, and conclusions are
given in Sec.~\ref{sec:Concl}. In App.~\ref{app:DiagRules},
diagrammatic rules of iPT used in this paper are briefly summarized.

\section{\label{sec:PSiPT}
The one-loop power spectra in the integrated perturbation theory
}

In this first section, the formalism of iPT \cite{mat11} is briefly
reviewed (without proofs), and formal expressions of power spectra in
the one-loop approximation are derived.

\subsection{Fundamental equations of the integrated perturbation theory
\label{subsec:}
}

In evaluating the power spectra in iPT, a concept of multi-point
propagator \cite{mat95,BCS08,BCS10} is useful. The $(n+1)$-point
propagator $\varGamma_X^{(n)}$ of any biased objects, which are
labeled by $X$ in general, is defined by \cite{mat11}
\begin{equation}
  \left\langle
      \frac{\delta^n \delta_X(\bm{k})}
      {\delta \delta_{\rm L}(\bm{k}_1) \cdots
       \delta \delta_{\rm L}(\bm{k}_n)}
 \right\rangle =
 (2\pi)^{3-3n}\delta_{\rm D}^3(\bm{k} - \bm{k}_{1\cdots n})
 \varGamma^{(n)}_X(\bm{k}_1,\ldots,\bm{k}_n),
\label{eq:1-1}
\end{equation}
where $\delta_X(\bm{k})$ is the Fourier transform of the number
density contrast of biased objects in Eulerian space, $\delta_{\rm
  L}(\bm{k})$ is the Fourier transform of linear density contrast,
$\delta_{\rm D}^3$ is the Dirac's delta function in three-dimensions,
and we adopt a notation
\begin{equation}
  \bm{k}_{1\cdots n} = \bm{k}_1 + \cdots + \bm{k}_n,
\label{eq:1-2}
\end{equation}
throughout this paper. The left-hand side of Eq.~(\ref{eq:1-1}) is an
ensemble average of $n$th-order functional derivative. The number
density field is considered as a functional of the initial density
field. In the basic framework of iPT, the biased objects can be any
astronomical objects which are observed as tracers of the underlying
density field in the universe.

The method how to evaluate multi-point propagators of biased objects
in the framework of iPT is detailed in Ref.~\cite{mat11}. In the most
general form of iPT formalism, both Eulerian and Lagrangian pictures
of dynamical evolutions can be dealt with, and both pictures give
equivalent predictions for observables. The models of halo bias fall
into the category of Lagrangian bias, i.e., the number density field
of halos is related to the mass density field in Lagrangian space. In
such a case, the Lagrangian picture is a natural way to describe
evolutions of halo number density field. In the models of Lagrangian
bias, the renormalized bias functions \cite{mat11} are the key
elements in iPT, which are defined by
\begin{equation}
    c^{(n)}_X(\bm{k}_1,\ldots,\bm{k}_n) = 
    (2\pi)^{3n} \int \frac{d^3k}{(2\pi)^3}
    \left\langle
        \frac{\delta^n\delta_X^{\rm L}(\bm{k})}
        {\delta\delta_{\rm L}(\bm{k}_1)\cdots\delta\delta_{\rm
            L}(\bm{k}_n)}
    \right\rangle,
\label{eq:1-3}
\end{equation}
where $\delta_X^{\rm L}(\bm{k})$ is the Fourier transform of halo
number density contrast in Lagrangian space. We allow the bias to be
nonlocal in Lagrangian space. In fact, the halo bias is not purely
local even in Lagrangian space \cite{mat12}. For a mass density field,
the Lagrangian number density contrast $\delta^\mathrm{L}_X$ is
identically zero, and the bias functions are identically zero,
$c^{(n)}_X = 0$ for all orders $n=1,2,\ldots$.

Assuming statistical homogeneity in Lagrangian space, the renormalized
bias functions in Eq.~(\ref{eq:1-3}) is equivalently defined by
\cite{mat12}
\begin{equation}
  \left\langle
      \frac{\delta^n \delta^{\rm L}_X(\bm{k})}
      {\delta\delta_{\rm L}(\bm{k}_1)
        \cdots\delta\delta_{\rm L}(\bm{k}_n)}
  \right\rangle = 
  (2\pi)^{3-3n}\delta_{\rm D}^3(\bm{k}-\bm{k}_{1\cdots n})
  c^{(n)}_X(\bm{k}_1,\ldots,\bm{k}_n).
\label{eq:1-4}
\end{equation}
The similarity of this equation with Eq.~(\ref{eq:1-1}) is apparent in
this form. The information on dynamics of bias in Lagrangian space is
encoded in the set of renormalized bias functions. Assuming
statistical isotropy in Lagrangian space, the renormalized bias
functions $c_X^{(n)}(\bm{k}_1,\ldots,\bm{k}_n)$ depend only on
magnitudes $k_1,\ldots,k_n$ and relative angles
$\hat{\bm{k}}_i\cdot\hat{\bm{k}}_j$ ($i>j$) of wavevectors.

Applying the vertex resummation of iPT, the multi-point propagators
of biased objects $X$ are given by a form,
\begin{equation}
  \varGamma^{(n)}_X(\bm{k}_1,\ldots,\bm{k}_n)
  = \varPi(\bm{k}_{1\cdots n})
  \hat{\varGamma}^{(n)}_X(\bm{k}_1,\ldots,\bm{k}_n),
\label{eq:1-5}
\end{equation}
where
\begin{equation}
  \varPi(\bm{k})
  = \left\langle e^{-i\bm{k}\cdot\bm{\varPsi}} \right\rangle
  = \exp\left[
      \sum_{n=2}^\infty 
      \frac{(-i)^n}{n!}
      \left\langle (\bm{k}\cdot\bm{\varPsi})^n \right\rangle_{\rm c}
  \right],
\label{eq:1-6}
\end{equation}
is the vertex resummation factor in terms of the displacement field
$\bm{\varPsi}$, and $\langle\cdots\rangle_{\rm c}$ indicates the
connected part of ensemble average. The displacement fields
$\bm{\varPsi}(\bm{q})$ are the fundamental variables in LPT, where
$\bm{q}$ is the Lagrangian coordinates and the Eulerian coordinates
are given by $\bm{x} = \bm{q} + \bm{\varPsi}(\bm{q})$. The cumulant
expansion theorem is used in the second equality of
Eq.~(\ref{eq:1-6}). Cumulants of the displacement fields with odd
number vanish from the parity symmetry, thus the summation in the
exponent of Eq.~(\ref{eq:1-6}) is actually taken over
$n=2,4,6,\ldots$. The normalized multi-point propagators of the biased
objects, $\hat{\varGamma}^{(n)}_X$, are naturally predicted in the
framework of iPT.

In the one-loop approximation of iPT, the vertex resummation factor is
given by
\begin{equation}
  \varPi(\bm{k}) =
  \exp\left\{
    -\frac12 \int\frac{d^3p}{(2\pi)^3}
    \left[\bm{k}\cdot\bm{L}^{(1)}(\bm{p})\right]^2
    P_{\rm L}(p)
  \right\},
\label{eq:1-7}
\end{equation}
and the normalized two-point propagator is given by
\begin{multline}
  \hat{\varGamma}_X^{(1)}(\bm{k})
  = c_X^{(1)}(k) + \bm{k}\cdot\bm{L}^{(1)}(\bm{k})
\\
  + \int\frac{d^3p}{(2\pi)^3} P_{\rm L}(p)
  \biggl\{
      c_X^{(2)}(\bm{k},\bm{p})
      \left[\bm{k}\cdot\bm{L}^{(1)}(-\bm{p})\right]
\\
      +\,c_X^{(1)}(p)
      \left[\bm{k}\cdot\bm{L}^{(1)}(-\bm{p})\right]
      \left[\bm{k}\cdot\bm{L}^{(1)}(\bm{k})\right]
\\
      + \frac12
      \bm{k}\cdot\bm{L}^{(3)}(\bm{k},\bm{p},-\bm{p})
\\
      +\,c_X^{(1)}(p)
      \left[\bm{k}\cdot\bm{L}^{(2)}(\bm{k},-\bm{p})\right]
\\
      + \left[\bm{k}\cdot\bm{L}^{(1)}(\bm{p})\right]
      \left[\bm{k}\cdot\bm{L}^{(2)}(\bm{k},-\bm{p})\right]
  \biggr\},
\label{eq:1-8}
\end{multline}
where $\bm{L}^{(n)}$ is the $n$th-order displacement kernel in LPT.

\begin{figure}[t]
\begin{center}
\includegraphics[width=20pc]{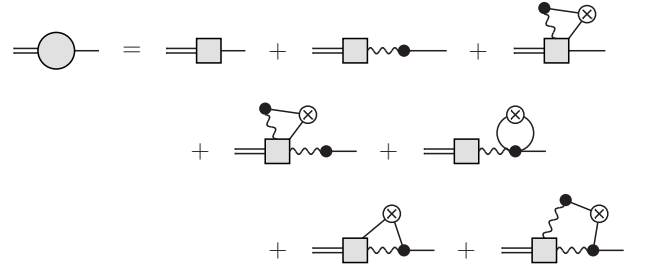}
\caption{\label{fig:PropOne} The diagrammatic representation of the
  two-point propagator with partially resummed vertex up to one-loop
  contributions. }
\end{center}
\end{figure}

Each term in Eq.~(\ref{eq:1-8}) respectively corresponds to each
diagram of Fig.~\ref{fig:PropOne} in the same order.
Diagrammatic rules in iPT \cite{mat11} with the Lagrangian picture,
which are explained in App.~\ref{app:DiagRules}, are applied in the
correspondence. The normalized two-point propagator of mass density
field, $\hat{\varGamma}^{(1)}_{\rm m}$, is obtained by putting
$c^{(n)}_X = 0$ in Eq.~(\ref{eq:1-8}).

The perturbative expansion of the displacement field in Fourier space,
$\tilde{\bm{\varPsi}}(\bm{k})$ is given by
\begin{equation}
  \tilde{\bm{\varPsi}}(\bm{k}) =
  \sum_{n=1}^\infty \frac{i}{n!}
  \int_{\bm{k}_{1\cdots n}=\bm{k}}
  \bm{L}^{(n)}(\bm{k}_1,\ldots,\bm{k}_n)
  \delta_{\rm L}(\bm{k}_1)\cdots\delta_{\rm L}(\bm{k}_n),
\label{eq:1-9}
\end{equation}
where we adopt a notation,
\begin{equation}
  \int_{\bm{k}_{1\cdots n}=\bm{k}}\cdots
  =
  \int \frac{d^3k_1}{(2\pi)^3} \cdots \frac{d^3k_n}{(2\pi)^3}
  (2\pi)^3\delta_{\rm D}^3\left(\bm{k}-\bm{k}_{1\cdots n}\right)
  \cdots.
\label{eq:1-10}
\end{equation}
Such notation as Eq.~(\ref{eq:1-10}) is commonly used throughout this
paper.

In real space, the kernels of LPT in the standard theory of gravity
(in the Newtonian limit) are given by \cite{cat95}
\begin{align}
& \bm{L}^{(1)}(\bm{k}) = \frac{\bm{k}}{k^2},
\label{eq:1-11a}\\
& \bm{L}^{(2)}(\bm{k}_1,\bm{k}_2)
  =\frac37 \frac{\bm{k}_{12}}{{k_{12}}^2}
  \left[1 - \left(\frac{\bm{k}_1 \cdot \bm{k}_2}{k_1 k_2}\right)^2\right],
\label{eq:1-11b}\\
&  \bm{L}^{(3)}(\bm{k}_1,\bm{k}_2,\bm{k}_3) =
  \frac13
  \left[\bm{L}^{(3{\rm a})}(\bm{k}_1,\bm{k}_2,\bm{k}_3) + {\rm perm.}\right];
\label{eq:1-11c}\\
& \bm{L}^{(3{\rm a})}(\bm{k}_1,\bm{k}_2,\bm{k}_3)
\nonumber\\
& \quad
  = \frac{\bm{k}_{123}}{{k_{123}}^2}
  \left\{
      \frac57
      \left[1 - \left(\frac{\bm{k}_1 \cdot \bm{k}_2}{k_1 k_2}\right)^2\right]
      \left[1 - \left(\frac{\bm{k}_{12} \cdot \bm{k}_3}
          {{k}_{12} k_3}\right)^2\right]
  \right.
\nonumber\\
& \qquad\quad
  \left.
  - \frac13
  \left[
      1 - 3\left(\frac{\bm{k}_1 \cdot \bm{k}_2}{k_1 k_2}\right)^2
      +\, 2 \frac{(\bm{k}_1 \cdot \bm{k}_2)(\bm{k}_2 \cdot \bm{k}_3)
        (\bm{k}_3 \cdot \bm{k}_1)}{{k_1}^2 {k_2}^2 {k_3}^2}
  \right]\right\}
\nonumber\\
& \qquad
  + \frac{\bm{k}_{123}}{{k_{123}}^2}
    \times \bm{T}(\bm{k}_1,\bm{k}_2,\bm{k}_3),
\label{eq:1-11d}
\end{align}
where a vector function $\bm{T}$ represents a transverse part whose
explicit expression will not be used in this paper. Complete
expressions of the displacement kernels of LPT up to 4th order,
including transverse parts are given in, e.g., Ref.~\cite{RB12,tat13}.
Eqs.~(\ref{eq:1-7}) and (\ref{eq:1-8}) remain valid even when the
non-standard theory of gravity is assumed as long as the appropriate
form of kernels $\bm{L}_n$ in such a theory is used.

One of the benefits in the Lagrangian picture is that redshift-space
distortions are relatively easy to be incorporated in the theory. A
displacement kernel in redshift space $\bm{L}^{{\rm s}(n)}$ is simply
related to the kernel in real space at the same order by a linear
mapping \cite{mat08a},
\begin{equation}
  \bm{L}^{(n)} \rightarrow \bm{L}^{{\rm s}(n)} = \bm{L}^{(n)} +
  nf\left(\hat{\bm{z}}\cdot\bm{L}^{(n)}\right)\hat{\bm{z}},
\label{eq:1-12}
\end{equation}
where $f=d\ln D/d\ln a = \dot{D}/HD$ is the linear growth rate, $D(t)$
is the linear growth factor, $a(t)$ is the scale factor, and $H(t) =
\dot{a}/a$ is the time-dependent Hubble parameter. The
distant-observer approximation is assumed in redshift space, and the
unit vector $\hat{\bm{z}}$ denotes the line-of-sight direction.
Strictly speaking, the mapping of Eq.~(\ref{eq:1-12}) is exact only in
the Einstein-de~Sitter universe. However, this mapping is a good
approximation in general cosmology. The expressions of
Eqs.~(\ref{eq:1-7}) and (\ref{eq:1-8}) apply as well in redshift space
when the displacement kernels in redshift space $\bm{L}^{{\rm s}(n)}$
are used instead of the real-space counterparts $\bm{L}^{(n)}$.

\begin{figure}[t]
\begin{center}
\includegraphics[width=20pc]{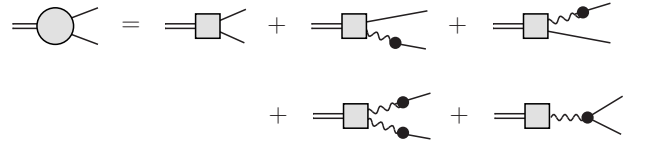}
\caption{\label{fig:PropTwo} The diagrammatic representation of the
  three-point propagator with partially resummed vertex at the
  tree-level contribution.}
\end{center}
\end{figure}

The three-point propagator at the tree-level approximation in iPT is
given by
\begin{multline}
  \hat{\varGamma}^{(2)}_X(\bm{k}_1,\bm{k}_2) = 
  c^{(2)}_X(\bm{k}_1,\bm{k}_2)
  + c^{(1)}_X(\bm{k}_1) \left[\bm{k}\cdot\bm{L}^{(1)}(\bm{k}_2)\right]
\\
  + c^{(1)}_X(\bm{k}_2) \left[\bm{k}\cdot\bm{L}^{(1)}(\bm{k}_1)\right]
  + \left[\bm{k}\cdot\bm{L}^{(1)}(\bm{k}_1)\right]
    \left[\bm{k}\cdot\bm{L}^{(1)}(\bm{k}_2)\right]
\\
  + \bm{k}\cdot\bm{L}^{(2)}(\bm{k}_1,\bm{k}_2), 
\label{eq:1-13}
\end{multline}
where each term respectively corresponds to each diagram of
Fig.~\ref{fig:PropTwo} in the same order.
When the mapping of Eq.~(\ref{eq:1-12}) is applied to every
displacement kernels in Eq.~(\ref{eq:1-13}), the expression of
three-point propagator in redshift space is obtained. The three-point
propagator of mass, $\varGamma^{(2)}_{\rm m}$, is given by just
substituting $c^{(n)}_X=0$ in Eq.~(\ref{eq:1-13}).

\begin{figure}[t]
\begin{center}
\includegraphics[width=20pc]{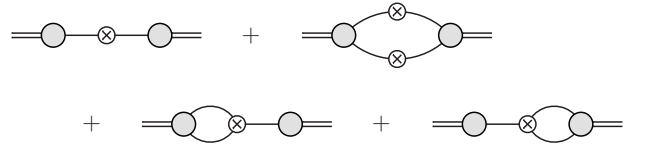}
\caption{\label{fig:HaloPS} The diagrammatic representation of the
  power spectrum up to one-loop approximation. }
\end{center}
\end{figure}

In terms of the multi-point propagators, the power spectrum of biased
objects, up to the one-loop approximation, is given by
\begin{multline}
  P_X(\bm{k}) = 
  \varPi^2(k)
  \Biggl\{
  \left[\hat{\varGamma}^{(1)}_X(\bm{k})\right]^2 P_{\rm L}(k)
\\
  + \frac12 \int_{\bm{k}_{12}=\bm{k}}
  \left[\hat{\varGamma}^{(2)}_X(\bm{k}_1,\bm{k}_2)\right]^2
  P_{\rm L}(k_1) P_{\rm L}(k_2)
\\
  + \hat{\varGamma}^{(1)}_X(\bm{k})
  \int_{\bm{k}_{12}=\bm{k}}
  \hat{\varGamma}^{(2)}_X(\bm{k}_1,\bm{k}_2)
  B_{\rm L}(k,k_1,k_2)
  \Biggr\},
\label{eq:1-14}
\end{multline}
where $P_{\rm L}(k)$ and $B_{\rm L}(k,k_1,k_2)$ are
the linear power spectrum and the linear bispectrum, respectively.
The diagrammatic representations of Eq.~(\ref{eq:1-14}) are shown in
Fig.~\ref{fig:HaloPS}.
Crossed circles correspond to the linear power spectrum or the linear
bispectrum, depending on number of lines attached to them. The first
two terms in Eq.~(\ref{eq:1-14}) corresponds to the first two diagrams
in Fig.~\ref{fig:HaloPS}. The last two diagrams in
Fig.~\ref{fig:HaloPS} are contributions from the primordial
non-Gaussianity. The two diagrams give the same contribution because
of the parity symmetry, and the sum of the two diagrams corresponds
to the last term in Eq.~(\ref{eq:1-14}).

The matter power spectrum $P_{\rm m}(\bm{k})$ is simply given by
replacing $\varGamma^{(n)}_X$ by $\varGamma^{(n)}_{\rm m}$ in
Eq.~(\ref{eq:1-14}), or equivalently, setting $c^{(n)}_X = 0$ for
every $n\geq 1$. The cross power spectrum between two types of
objects, $X$ and $Y$, is similarly obtained as
\begin{multline}
  P_{XY}(\bm{k}) = 
  \varPi^2(k)
  \Biggl\{
  \hat{\varGamma}^{(1)}_X(\bm{k})
  \hat{\varGamma}^{(1)}_Y(\bm{k}) P_{\rm L}(k)
\\
  + \frac12 \int_{\bm{k}_{12}=\bm{k}}
  \hat{\varGamma}^{(2)}_X(\bm{k}_1,\bm{k}_2)
  \hat{\varGamma}^{(2)}_Y(\bm{k}_1,\bm{k}_2)
  P_{\rm L}(k_1) P_{\rm L}(k_2)
\\
  + \frac12 \hat{\varGamma}^{(1)}_X(\bm{k})
  \int_{\bm{k}_{12}=\bm{k}}
  \hat{\varGamma}^{(2)}_Y(\bm{k}_1,\bm{k}_2)
  B_{\rm L}(k,k_1,k_2)
\\
  + \frac12 \hat{\varGamma}^{(1)}_Y(\bm{k})
  \int_{\bm{k}_{12}=\bm{k}}
  \hat{\varGamma}^{(2)}_X(\bm{k}_1,\bm{k}_2)
  B_{\rm L}(k,k_1,k_2)
    \Biggr\}.
\label{eq:1-15}
\end{multline}
The diagrams for the above equations are similar to the ones in
Fig.~\ref{fig:HaloPS}, where the left and right multi-point
propagators correspond to those of $X$ and $Y$, respectively. When
$X=Y$, Eq.~(\ref{eq:1-15}) apparently reduces to Eq.~(\ref{eq:1-14}).

The predictions of biased power spectra in the one-loop approximation
of iPT are given by Eq.~(\ref{eq:1-14}) for the auto power spectrum,
and by Eq.~(\ref{eq:1-15}) for the cross power spectrum. Once a model
of the renormalized bias functions $c^{(n)}_X$ is given, it is
straightforward to numerically evaluate those equations. The above
results are general and do not depend on bias models. Any bias model
can be incorporated in the expression of iPT through the renormalized
bias functions. In the next subsection, we explain a simple model of
the renormalized bias function based on the halo approach.

\subsection{Renormalized bias functions in a simple model of halo approach
\label{subsec:HaloBias}
}

The renormalized bias functions $c^{(n)}_X$ are not specified in the
general framework of iPT. Precise modeling of bias is a nontrivial
problem, depending on what kind of biased tracers are considered. In
this subsection, we consider a simple model of halo bias as an
example. The expressions of renormalized bias functions in a simple
model of the halo approach are recently derived in Ref.~\cite{mat12}.
We summarize the consequences of this model below. It should be
emphasized that the general framework of iPT does not depend on this
specific model of bias.

Without resorting to approximations such as the peak-background split,
the halo bias is shown to be nonlocal even in Lagrangian space. As a
result, the renormalized bias functions have nontrivial scale
dependencies. For the halos of mass $M$, the renormalized bias
functions are given by \cite{mat12}
\begin{multline}
  c^{(n)}_M(\bm{k}_1,\ldots,\bm{k}_n) =
  b^{\rm L}_n(M) W(k_1R) \cdots W(k_nR)
\\
  + \frac{A_{n-1}(M)}{{\delta_{\rm c}}^n}
  \frac{d}{d\ln\sigma_M}
    \left[W(k_1R) \cdots W(k_nR)\right],
\label{eq:1-20}
\end{multline}
where $\delta_{\rm c}=3(3\pi/2)^{2/3}/5\simeq 1.686$ is the critical
overdensity for spherical collapse and $W(kR)$ is a window function.
In a usual halo approach, the window function is chosen to be a
top-hat type in configuration space, which corresponds to
\begin{equation}
   W(x) = \frac{3\sin x - 3x \cos x}{x^3},
\label{eq:1-23}
\end{equation}
in Fourier space. In this case, the Lagrangian radius $R$ is naturally
related to the mass $M$ of halo by
\begin{equation}
  M = \frac43 \pi \bar{\rho}_0 R^3,
\label{eq:1-21}
\end{equation}
where $\bar{\rho}_0$ is the mean density of mass at the present time,
or
\begin{equation}
  R = \left[
      \frac{M}{1.163\times 10^{12} h^{-1} M_\odot \varOmega_{\rm m0}}
      \right]^{1/3}\,h^{-1}\rm{Mpc},
\label{eq:1-22}
\end{equation}
where $M_\odot =1.989\times 10^{30}\,{\rm kg}$ is the solar mass,
$\varOmega_{\rm m0}$ is the density parameter of mass at the present
time, and $h=H_0/(100\, \rm{km\,s^{-1}\,Mpc^{-1}})$ is the normalized
Hubble constant. 

Empirically, one can also use other types of window function. Direct
evaluations of the renormalized bias functions suggest that Gaussian
window function $W(x) = e^{-k^2 R^2/2}$ gives better fit \cite{NMT14}.
In the latter case, the relation between the smoothing radius $R$ and
mass $M$ is not trivial and should also be empirically modified from
the relation of Eq.~(\ref{eq:1-22}). However, the shapes of one-loop
power spectrum on large scales are not sensitive to the choice of
window function.

The variance of density fluctuations on the mass scale $M$ is defined
by
\begin{equation}
  {\sigma_M}^2 = \int \frac{d^3k}{(2\pi)^3} W^2(kR) P_{\rm L}(k).
\label{eq:1-24}
\end{equation}
The radius $R$ is considered as a function of $\sigma_M$ through
Eq.~(\ref{eq:1-20}). The functions $A_n(M)$ are defined by
\begin{equation}
  A_n(M) \equiv
  \sum_{j=0}^n \frac{n!}{j!}{\delta_{\rm c}}^j\, b^{\rm L}_j(M),
\label{eq:1-25}
\end{equation}
where $b^{\rm L}_n$ is the scale-independent Lagrangian bias parameter
of $n$th-order. For example, first three functions are given by
\begin{equation}
  A_0 = 1, \quad
  A_1 = 1 + \delta_{\rm c} b^{\rm L}_1, \quad
  A_2 = 2 + 2\delta_{\rm c} b^{\rm L}_1 + {\delta_{\rm c}}^2 b^{\rm L}_2.
\label{eq:1-25-1}
\end{equation}
When the halo mass function $n(M)$ takes a universal form
\begin{equation}
  n(M)dM = \frac{\bar{\rho}_0}{M} f_{\rm MF}(\nu) \frac{d\nu}{\nu},
\label{eq:1-26}
\end{equation}
where $\nu = \delta_{\rm c}/\sigma_M$, the Lagrangian bias
parameters are given by
\begin{equation}
  b^{\rm L}_n(M) =
  \left(\frac{-1}{\sigma_M}\right)^n
      \frac{f_{\rm MF}^{(n)}(\nu)}{f_{\rm MF}(\nu)},
\label{eq:1-27}
\end{equation}
where $f_{\rm MF}^{(n)} = d^nf_{\rm MF}/d\nu^n$.

Once the model of the mass function $f_{\rm MF}(\nu)$ is given, the
scale-independent bias parameters $b^{\rm L}_n(M)$ and the functions
$A_n(M)$ are uniquely given by Eqs.~(\ref{eq:1-27}) and
(\ref{eq:1-25}). In Table~\ref{tab:BiasFns}, those functions are
summarized for popular models of mass function, i.e., the
Press-Schechter (PS) mass function \cite{PS74}, the Sheth-Tormen (ST)
mass function \cite{ST99}, Warren {\it et al.} (W+) mass function
\cite{War06}.
\begin{table*}
\begin{tabular}{c|c|c|c}
  {}
  & PS
  & ST
  & W+, MICE ($\sigma=\delta_{\rm c}/\nu$)
  \\ \hline
  $f_{\rm MF}(\nu)$
  & $\displaystyle \sqrt{\frac{2}{\pi}}\,\nu e^{-\nu^2/2}$ 
  & $\displaystyle A(p) \sqrt{\frac{2}{\pi}}
  \left[1 + \frac{1}{(q\nu^2)^p}\right]\sqrt{q}\,\nu e^{-q\nu^2/2}$
  & $\displaystyle A \left(\sigma^{-a} + b\right) e^{-c/\sigma^2}$
  \\[.7pc]
  $b^{\rm L}_1(M)$    
  & $\displaystyle \frac{\nu^2 - 1}{\delta_{\rm c}}$
  & $\displaystyle \frac{1}{\delta_{\rm c}} \left[q\nu^2 - 1 +
    \frac{2p}{1 + (q\nu^2)^p}\right]$
  & $\displaystyle \frac{1}{\delta_{\rm c}}
  \left(\frac{2c}{\sigma^2} - \frac{a}{1+b\sigma^a}\right)$
  \\[.7pc]
  $b^{\rm L}_2(M)$
  & $\displaystyle \frac{\nu^4 - 3\nu^2}{{\delta_{\rm c}}^2}$
  & $\displaystyle \frac{1}{{\delta_{\rm c}}^2} \left[q^2\nu^4 - 3
    q\nu^2 + \frac{2p(2q\nu^2 + 2p -1)}{1 + (q\nu^2)^p}\right]$
  & $\displaystyle \frac{1}{{\delta_{\rm c}}^2}
  \left[ \frac{4c^2}{\sigma^4} - \frac{2c}{\sigma^2}
    - \frac{a\left(4c/\sigma^2 - a + 1\right)}{1+b\sigma^a}
  \right]$
  \\[.7pc]
  $A_1(M)$
  & $\nu^2$
  & $\displaystyle q\nu^2 + \frac{2p}{1 + (q\nu^2)^p} $
  & $\displaystyle \frac{2c}{\sigma^2} + 1 - \frac{a}{1+b\sigma^a}$
  \\[.7pc]
  $A_2(M)$
  & $\nu^2(\nu^2-1)$
  & $\displaystyle q\nu^2 (q\nu^2 -1) + \frac{2p(2q\nu^2 + 2p +
    1)}{1 + (q\nu^2)^p}$
  & $\displaystyle \frac{4c^2}{\sigma^4}
  + \frac{2c}{\sigma^2} + 2
  - \frac{a\left(4c/\sigma^2 - a + 3\right)}{1+b\sigma^a}$
  \\[.7pc]\hline
  Parameters
  & -
  & $ \begin{array}{c}
    A(p) = [1+\pi^{-1/2}2^{-p}\varGamma(1/2-p)]^{-1} \\
    p=0.3 \\ q=0.707 \end{array}$
  & $ \begin{array}{l} \mbox{W+}:\\ A = 0.7234 \\
      a = 1.625 \\ b=0.2538 \\ c=1.1982 \end{array}$
    $ \begin{array}{l} \mbox{MICE}: \\
      A(z) = 0.58(1+z)^{-0.13}\\ a(z) = 1.37(1+z)^{-0.15} \\ b(z) =
        0.3(1+z)^{-0.084} \\ c(z)=1.036(1+z)^{-0.024} \end{array}$
\end{tabular}
\caption{Functions $b^{\rm L}_n(M)$, $A_n(M)$ derived from several
  models of mass function.
\label{tab:BiasFns}}
\end{table*}

In the simplest PS mass function, it is interesting to note that
general expressions of the parameters for all orders can be derived
\cite{mat12}: $b^{\rm L}_n = \nu^{n-1}H_{n+1}(\nu)/{\delta_{\rm
    c}}^n$, $A_n = \nu^n H_n(\nu)$, where $H_n(\nu)$ are the Hermite
polynomials. The ST mass function gives a better fit to numerical
simulations of halos in cold-dark-matter type cosmologies with
Gaussian initial conditions. The values of parameters in
Table~\ref{tab:BiasFns} are $p=0.3$, $q=0.707$, and $A(p) =
[1+\pi^{-1/2}2^{-p}\varGamma(1/2-p)]^{-1}$ is the normalization
factor. When we put $p=0$, $q=1$, the ST mass function reduces to the
PS mass function. The W+ mass function is represented by a parameter
$\sigma = \delta_{\rm c}/\nu$, which is also a function of $M$, and
parameters are $A = 0.7234$, $a=1.625$, $b=0.2538$, $c=1.1982$. The
same functional form is applied to the Marenostrum Institut de
Ci\'encies de l'Espai (MICE) simulations in Ref.~\cite{Cro10},
allowing the parameters redshift-dependent. Their values are given by
$A(z) = 0.58(1+z)^{-0.13}$, $a(z) = 1.37(1+z)^{-0.15}$, $b(z) =
0.3(1+z)^{-0.084}$, $c(z)=1.036(1+z)^{-0.024}$. When the
redshift-dependent parameters are adopted, the W+ mass function is
sometimes referred to as the ``MICE mass function''. In the latter
case, the multiplicity function $f_{\rm MF}(\nu)$ explicitly depends
on the redshift, and the mass function is no longer 'universal'.

The nonlocal nature of the halo bias in Lagrangian space is encoded in
the second term in the RHS of Eq.~(\ref{eq:1-20}), since the simple
dependence on the window function of the first term appears even in
the local bias models through the smoothed mass density field. In the
large-scale limit, $k_1,k_2,\ldots,k_n \rightarrow 0$, the second term
in the RHS of Eq.~(\ref{eq:1-20}) disappears and the renormalized bias
functions reduce to scale-independent bias parameters, $c^{(n)}_M
\simeq b^{\rm L}_n(M)$. This property is consistent with the
peak-background split. However, the loop corrections in the iPT
involve integrations over the wavevectors of the renormalized bias
functions, and there is no reason to neglect the second term which
represents nonlocal nature of Lagrangian bias of halos.

The Eq.~(\ref{eq:1-20}) is shown to be equivalent to the following
expression \cite{mat12},
\begin{multline}
  c^{(n)}_M(\bm{k}_1,\ldots,\bm{k}_n) =
  \frac{A_n(M)}{{\delta_{\rm c}}^n} W(k_1R) \cdots W(k_nR)
\\
  + \frac{A_{n-1}(M)\, {\sigma_M}^n}{{\delta_{\rm c}}^n}
  \frac{d}{d\ln\sigma_M}
    \left[\frac{W(k_1R) \cdots W(k_nR)}{{\sigma_M}^n} \right].
\label{eq:1-28}
\end{multline}
For the PS mass function, there is an interesting relation, $A_n =
\nu^2 {\delta_{\rm c}}^{n-1}b^{\rm L}_{n-1}$, and in this case, the
renormalized bias function $c^{\rm L}_n$ is expressible by lower-order
parameters $b^{\rm L}_{n-1}$ and $b^{\rm L}_{n-2}$, which is a reason
why the scale-dependent bias in the presence of primordial
non-Gaussianity is approximately proportional to the first-order bias
parameter, $b^{\rm L}_1$ rather than the second-order one, $b^{\rm
  L}_2$ \cite{mat12}. However, this does not mean that $c^{(n)}_M$ is
independent on $b^{\rm L}_n$, because $b^{\rm L}_n$ can be expressible
by a linear combination of $b^{\rm L}_{n-1}$ and $b^{\rm L}_{n-2}$ in
the PS mass function.

In the expressions of renormalized bias functions,
Eqs.~(\ref{eq:1-20}) and (\ref{eq:1-28}), all the halos are assumed to
have the same mass, $M$. These expressions apply when the mass range
of halos in a given sample is sufficiently narrow. When the mass range
is finitely extended, the expressions should be replaced by
\cite{mat12}
\begin{equation}
  c^{(n)}_\phi(\bm{k}_1,\ldots,\bm{k}_n) = 
  \frac{\int dM\,\phi(M)\,n(M)\,c^{(n)}_M(\bm{k}_1,\ldots,\bm{k}_n)}
  {\int dM\,\phi(M)\,n(M)},
\label{eq:1-40}
\end{equation}
where $n(M)$ is the halo mass function of Eq.~(\ref{eq:1-26}),
$\phi(M)$ is a selection function of mass. For a simple example, when
the mass of halos are selected by a finite range $[M_1,M_2]$,
we have
\begin{equation}
  c^{(n)}_{[M_1,M_2]}(\bm{k}_1,\ldots,\bm{k}_n) = 
  \frac{\int_{M_1}^{M_2} dM\,n(M)\,c^{(n)}_M(\bm{k}_1,\ldots,\bm{k}_n)}
    {\int_{M_1}^{M_2} dM\,n(M)}.
\label{eq:1-41}
\end{equation}

\section{Explicit Formulas
\label{sec:ExplicitFormulas}
}

The auto power spectrum $P_X(\bm{k})$ of Eq.~(\ref{eq:1-14}) is a
special case of the cross power spectrum $P_{XY}(\bm{k})$ of
Eq.~(\ref{eq:1-15}) as the former is given by setting $X=Y$ in the
latter. It is general enough to give the formulas for the cross power
spectrum below. In the following, we decompose Eq.~(\ref{eq:1-15})
into the following form:
\begin{equation}
  P_{XY}(\bm{k}) =
  \varPi^2(\bm{k})
  \left[
      R_{XY}(\bm{k}) + Q_{XY}(\bm{k}) + S_{XY}(\bm{k})
  \right],
\label{eq:2-1}
\end{equation}
where $\varPi(\bm{k})$ is given by Eq.~(\ref{eq:1-7}) and
\begin{align}
  R_{XY}(\bm{k}) &= \hat{\varGamma}^{(1)}_X(\bm{k})
  \hat{\varGamma}^{(1)}_Y(\bm{k}) P_{\rm L}(k),
\label{eq:2-2a}\\
  Q_{XY}(\bm{k}) &=
  \frac12
  \int_{\bm{k}_{12}=\bm{k}}
  \hat{\varGamma}^{(2)}_X(\bm{k}_1,\bm{k}_2)
  \hat{\varGamma}^{(2)}_Y(\bm{k}_1,\bm{k}_2)
  P_{\rm L}(k_1) P_{\rm L}(k_2),
\label{eq:2-2b}\\
  S_{XY}(\bm{k}) &=
  \frac12
  \hat{\varGamma}^{(1)}_X(\bm{k})
  \int_{\bm{k}_{12}=\bm{k}}
  \hat{\varGamma}^{(2)}_Y(\bm{k}_1,\bm{k}_2)
  B_{\rm L}(k,k_1,k_2)
\nonumber\\
& \qquad
  + (X \leftrightarrow Y).
\label{eq:2-2c}
\end{align}
Three-dimensional integrals appeared in the above components of
Eqs.~(\ref{eq:2-2a})--(\ref{eq:2-2c}) can be reduced to
lower-dimensional integrals both in real space and in redshift space.
Such dimensional reductions of the integrals are useful for practical
calculations. The purpose of this section is to give explicit formulas
for the above components $\varPi$, $R_{XY}$, $Q_{XY}$, $S_{XY}$ in
terms of two-dimensional integrals at most. The results of this
section are applicable to any bias models, and do {\em not} depend on
specific forms of renormalized bias functions, e.g., those explained
in Sec.~\ref{subsec:HaloBias}.

\subsection{The power spectra in real space
\label{subsec:RealSpacePS}}

In real space, the power spectrum is independent on the direction of
wavevector $\bm{k}$, and thus the components above $\varPi(k)$,
$R_{XY}(k)$, $Q_{XY}(k)$, $S_{XY}(k)$ are also independent on the
direction. In this case, dimensional reductions of the integrals in
Eqs.~(\ref{eq:2-2a})--(\ref{eq:2-2c}) are not difficult, because of
the rotational symmetry. The vertex resummation factor
$\varPi(\bm{k})$ of Eq.~(\ref{eq:1-7}) is given by
\begin{equation}
  \varPi(k)
  = \exp\left[
       -\frac{k^2}{12\pi^2}\int dp\,P_{\rm L}(p)
     \right].
\label{eq:2-3}
\end{equation}
On small scales, this factor exponentially suppresses the power too
much, and such a behavior is not physical. This property is a good
indicator of which scales the perturbation theory should not be
applied. However, the resummation of the vertex factor is not
compulsory in the iPT. When the vertex factor is not resummed, one can
expand the factor as
\begin{equation}
  \varPi(k)
  = 1 -\frac{k^2}{12\pi^2}\int dp\,P_{\rm L}(p),
\label{eq:2-4}
\end{equation}
instead of Eq.~(\ref{eq:2-3}) in the case of one-loop perturbation
theory. In a quasi-linear regime, the resummed vertex factor of
Eq.~(\ref{eq:2-3}) gives better fit to $N$-body simulations in real
space \cite{mat08a, SM11}.

The expression of two-point propagator in Eq.~(\ref{eq:1-8}) is
straightforwardly obtained, substituting the Lagrangian kernels of
Eqs.~(\ref{eq:1-11a})--(\ref{eq:1-11d}). Taking the $z$-axis of
$\bm{p}$ as the direction of $\bm{k}$, integrations by the azimuthal
angle are trivial. Transforming the rest of integration variables as
$r = p/k$ and $x = \hat{\bm{p}}\cdot\hat{\bm{k}}$, we have two
equivalent expressions,
\begin{align}
  \hat{\varGamma}^{(1)}_X(k) &= 
  1 + c^{(1)}_X(k)
  + \frac{k^3}{4\pi^2} \int_0^\infty dr \int_{-1}^1 dx\,
    \hat{\cal R}_X(k,r,x) P_{\rm L}(kr)
\label{eq:2-5a}\\
  &= 1 + c^{(1)}_X(k)
  + \frac{k^3}{4\pi^2} \int_0^\infty dr\, \tilde{\cal R}_X(k,r)
  P_{\rm L}(kr), 
\label{eq:2-5b}
\end{align}
where
\begin{multline}
  \hat{\cal R}_X(k,r,x) = \frac{5}{21} \frac{r^2(1-x^2)^2}{1+r^2-2rx}
\\
  + \frac37 \frac{(1-rx)(1-x^2)}{1+r^2-2rx}
    \left[ rx + r^2c^{(1)}_X(kr) \right]
\\
  - rx\,c^{(2)}_X(k,kr;x),
\label{eq:2-6}
\end{multline}
and
\begin{multline}
  \tilde{\cal R}_X(k,r) = 
  \frac{6 + 5 r^2 + 50 r^4 - 21 r^6}{252 r^2}
\\
  + \frac{(1 - r^2)^3 (2+7r^2)}{168 r^3}
  \ln\left|\frac{1-r}{1+r}\right|
\\
  + \left[
      \frac{3 + 8r^2 - 3r^4}{28} 
      + \frac{3(1-r^2)^3}{56r}  \ln\left|\frac{1-r}{1+r}\right|
    \right] c^{(1)}_X(kr)
\\
  - r \int_{-1}^1 dx\, x\,c^{(2)}_X(k,kr;x).
\label{eq:2-7}
\end{multline}
In the above expressions, rotationally invariant arguments for
$c^{(2)}_X$ are used, i.e.,
\begin{equation}
  c^{(2)}_X(\bm{k}_1,\bm{k}_2)
  = c^{(2)}_X(k_1,k_2;x),
\label{eq:2-8}
\end{equation}
where $x = \hat{\bm{k}}_1\cdot\hat{\bm{k}}_2$ is the direction
cosine between $\bm{k}_1$ and $\bm{k}_2$. The second expression of
Eq.~(\ref{eq:2-5b}) is obtained by analytically integrating the
variable $x$ in the first expression of Eq.~(\ref{eq:2-5a}). Both
expressions are suitable for numerical evaluations. With the
expression of Eq.~(\ref{eq:2-5a}) or (\ref{eq:2-5b}), we have
\begin{equation}
  R_{XY}(k) =
  \hat{\varGamma}^{(1)}_X(k) \hat{\varGamma}^{(1)}_Y(k) P_{\rm L}(k).
\label{eq:2-9}
\end{equation}

Evaluating the convolution integrals in Eqs.~(\ref{eq:2-2b}) and
(\ref{eq:2-2c}) with the three-point propagator of Eq.~(\ref{eq:1-13})
is also straightforward in real space. Substituting the Lagrangian
kernels of Eqs.~(\ref{eq:1-11a}) and (\ref{eq:1-11b}) into
Eq.~(\ref{eq:1-13}), and transforming the integration variables as
$r=k_1/k$, $x=\hat{\bm{k}}\cdot\hat{\bm{k}_1}$, we have
\begin{multline}
  Q_{XY}(k) =
  \frac{k^3}{8\pi^2} \int_0^\infty dr \int_{-1}^1 dx\,r^2
  \hat{\varGamma}^{(2)}_X(k,r,x) \hat{\varGamma}^{(2)}_Y(k,r,x)
\\
  \times
  P_{\rm L}(kr) P_{\rm L}(ky)
\label{eq:2-10}
\end{multline}
and
\begin{multline}
  S_{XY}(k) = 
  \frac{k^3}{8\pi^2}
  \hat{\varGamma}^{(1)}_X(k)
  \int_0^\infty dr \int_{-1}^1 dx\,r^2
  \hat{\varGamma}^{(2)}_Y(k,r,x)
\\ \times
  B_{\rm L}(k,kr,ky)
  + (X\leftrightarrow Y),
\label{eq:2-11}
\end{multline}
where
\begin{equation}
  y = \sqrt{1+r^2-2rx},
\label{eq:2-12}
\end{equation}
and
\begin{multline}
  \hat{\varGamma}^{(2)}_X(k,r,x) = 
  -\frac47\,\frac{1-x^2}{y^2}
  + \frac{x}{r}\left[1+c^{(1)}_X(ky)\right]
\\
  + \frac{1-rx}{y^2}\left[1+ c^{(1)}_X(kr)\right]
  + c^{(2)}_X(kr,ky;x),
\label{eq:2-13}
\end{multline}
The factor $\hat{\varGamma}^{(2)}_Y(k,r,x)$ is similarly given by
substituting $X\rightarrow Y$ in Eq.~(\ref{eq:2-13}). The function
$\hat{\varGamma}^{(2)}_X(k,r,x)$ is just the normalized three-point
propagator $\hat{\varGamma}^{(2)}_X(\bm{k}_1,\bm{k}-\bm{k}_1)$ as a
function of transformed variables.

All the necessary components to calculate the power spectrum of
Eq.~(\ref{eq:2-1}) in real space,
\begin{equation}
  P_{XY}(k) =
  \varPi^2(k)
  \left[
      R_{XY}(k) + Q_{XY}(k) + S_{XY}(k)
  \right],
\label{eq:2-14}
\end{equation}
are given above, i.e., Eqs.~(\ref{eq:2-3}) [or (\ref{eq:2-4})],
(\ref{eq:2-9}), (\ref{eq:2-10}) and (\ref{eq:2-11}). Numerical
integrations of Eqs.~(\ref{eq:2-5b}) [or (\ref{eq:2-5a})],
(\ref{eq:2-10}) and (\ref{eq:2-11}) are not difficult, once the model
of renormalized bias functions $c^{(n)}_X$ and primordial spectra
$P_{\rm L}(k)$, $B_{\rm L}(k_1,k_2,k_3)$ are given. The last factor
$S_{XY}(k)$ is absent in the case of Gaussian initial conditions.

\subsection{Kernel integrals
\label{subsec:KernelIntegrals}}

Evaluations of power spectra in redshift space are more tedious than
those in real space. The reason is that the power spectra depend on
the lines-of-sight direction in redshift space. One cannot arbitrary
choose the direction of $z$-axis in the three-dimensional integrations
of Eqs.~(\ref{eq:1-8}) and (\ref{eq:2-2a})--(\ref{eq:2-2c}), because
the rotational symmetry is not met. Even in such cases, an axial
symmetry around the lines of sight remains, and the three-dimensional
integrations can be reduced to two- or one-dimensional integrations as
shown below. All the necessary techniques for such reductions are the
same with those presented in Refs.~\cite{mat08a, mat08b}, making use
of rotational covariance. We summarize useful formulas for the
reduction in this subsection. We assume the standard theory of gravity
in the formula below, although the same technique may be applicable to
other theories such as the modified gravity, etc.

The first set of formulas is related to the two-point propagator
$\varGamma^{(1)}_X$ of Eq.~(\ref{eq:1-8}). The results are
summarized in Table~\ref{tab:Rformulas}. The integrals of a form,
\begin{equation}
  \displaystyle\vphantom{\frac{\int}{\int}}
  \int \frac{d^3p}{(2\pi)^3}
  {\cal F}(\bm{k},\bm{p}) P_{\rm L}(p),
\label{eq:2-20}
\end{equation}
where ${\cal F}(\bm{k},\bm{p})$ consists of LPT kernels
$\bm{L}^{(n)}$ and renormalized bias functions $c^{(n)}_X$, are
reduced to one-dimensional integrals, $R^X_n(k)$. The explicit
formulas are given in Table~\ref{tab:Rformulas}.
\begin{table}
\begin{tabular}{c|c|c}
    $\displaystyle {\cal F}(\bm{k},\bm{p})$
    & $\displaystyle\vphantom{\frac{\int}{\int}}
    \int \frac{d^3p}{(2\pi)^3}
    {\cal F}(\bm{k},\bm{p}) P_{\rm L}(p)$
    & Diagram
    \\[.5pc] \hline\hline
    $\displaystyle \bm{L}^{(3)}(\bm{k},\bm{p},-\bm{p}) $
    & $\displaystyle\vphantom{\frac{\int}{\int}}
    \frac{10}{21} \frac{\bm{k}}{k^2} R_1(k)$
    &
    \includegraphics[height=1.8pc]{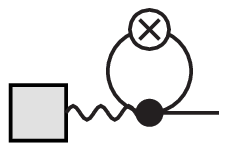}
    \\[.5pc]\hline
    $\displaystyle L^{(1)}_i(-\bm{p})L^{(2)}_j(\bm{k},\bm{p}) $
    & $\displaystyle\vphantom{\frac{\int}{\int}}
     \frac{3}{14} \frac{k_ik_j - k^2 \delta_{ij}}{k^4} R_1(k)
     + \frac{3}{7} \frac{k_ik_j}{k^4} R_2(k) $
    &
    \includegraphics[height=1.8pc]{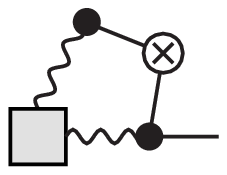}
    \\[.5pc]\hline
    $\displaystyle \bm{L}^{(2)}(\bm{k},\bm{p}) c^{(1)}_X(p) $
    & $\displaystyle\vphantom{\frac{\int}{\int}}
    \frac37 \frac{\bm{k}}{k^2} R_3^X(k) $
    &
    \includegraphics[height=1.8pc]{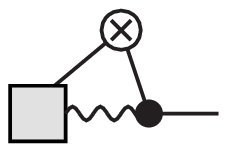}
    \\[.5pc]\hline
    $\displaystyle \bm{L}^{(1)}(-\bm{p}) c^{(2)}_X(\bm{k},\bm{p}) $
    & $\displaystyle\vphantom{\frac{\int}{\int}}
    - \frac{\bm{k}}{k^2} R_4^X(k) $
    &
    \includegraphics[height=1.8pc]{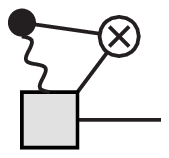}
\end{tabular}
\caption{Integral formulas for one-loop corrections, which are related
  to the two-point propagator. We denote $R_1(k) = R_1^X(k)$ and
  $R_2(k) = R_2^X(k)$, as these functions are independent on the bias.
  \label{tab:Rformulas}}
\end{table}
In this Table, we denote $R_1(k) = R_1^X(k)$ and $R_2(k) = R_2^X(k)$,
as these functions are independent on the bias. The functions
$R^X_n(k)$ are defined by three equivalent sets of equations,
\begin{align}
  R^X_n(k) &= \int\frac{d^3p}{(2\pi)^3} {\cal R}^X_n(\bm{k},\bm{p})
  P_{\rm L}(p)
\nonumber\\
  &= 
  \frac{k^3}{4\pi^2}
  \int_0^\infty dr \int_{-1}^1 dx\,
  \hat{\cal R}^X_n(r,x) P_{\rm L}(kr)
\nonumber\\
  &= 
  \frac{k^3}{4\pi^2}
  \int_0^\infty dr\, \tilde{\cal R}^X_n(r) P_{\rm L}(kr),
\label{eq:2-21}
\end{align}
where integrands ${\cal R}^X_n(\bm{k},\bm{p})$, $\hat{\cal
  R}^X_n(r,x)$, and $\tilde{\cal R}^X_n(r)$ are given in
Table~\ref{tab:Rfns}.
\begin{table*}
\begin{tabular}{c|c|c|c}
    $n$
    & ${\cal R}^X_n(\bm{k},\bm{p})$
    & $\hat{\cal R}^X_n(r,x)$
    & $\tilde{\cal R}^X_n(r)$
    \\[.3pc] \hline
    $1$
    & $\displaystyle\frac{k^2}{|\bm{k}-\bm{p}|^2} \left[1 -
        \left(\frac{\bm{k}\cdot\bm{p}}{kp}\right)^2\right]^2$
    & $\displaystyle\frac{r^2(1-x^2)^2}{1+r^2-2rx}$
    & $\displaystyle
    - \frac{(1+r^2)(3-14r^2+3r^4)}{24r^2}
    - \frac{(1-r^2)^4}{16r^3}\ln\left|\frac{1-r}{1+r}\right|$
    \\[.7pc]
    $2$
    & $\displaystyle \frac{\left(\bm{k}\cdot\bm{p}\right)
      \left[\bm{k}\cdot(\bm{k}-\bm{p})\right]}{p^2|\bm{k}-\bm{p}|^2}
    \left[1 - \left(\frac{\bm{k}\cdot\bm{p}}{kp}\right)^2\right]$
    & $\displaystyle  \frac{rx(1-rx)(1-x^2)}{1+r^2-2rx} $
    & $\displaystyle  \frac{(1-r^2)(3-2r^2+3r^4)}{24r^2}
    + \frac{(1-r^2)^3(1+r^2)}{16r^3}\ln\left|\frac{1-r}{1+r}\right|$
    \\[.7pc]
    $3$
    & $\displaystyle \frac{\bm{k}\cdot(\bm{k}-\bm{p})}
    {|\bm{k}-\bm{p}|^2}
    \left[1 - \left(\frac{\bm{k}\cdot\bm{p}}{kp}\right)^2\right]
    c^{(1)}_X(p) $
    & $\displaystyle  \frac{r^2(1-rx)(1-x^2)}{1+r^2-2rx} c^{(1)}_X(kr) $
    & $\displaystyle  \left[
        \frac{3+8r^2-3r^4}{12}
        + \frac{(1-r^2)^3}{8r}\ln\left|\frac{1-r}{1+r}\right|
    \right] c^{(1)}_X(kr) $
    \\[.7pc]
    $4$
    & $\displaystyle  \frac{\bm{k}\cdot\bm{p}}{p^2}
    c^{(2)}_X(\bm{k},\bm{p}) $
    & $\displaystyle  rx\, c^{(2)}_X(k,kr;x) $
    & $\displaystyle r \int_{-1}^1 dx\,
    x\, c^{(2)}_X(k,kr;x) $
    \\[.7pc]
\end{tabular}
\caption{Integrands for functions $R^X_n(k)$ of Eq.~(\ref{eq:2-21}).
\label{tab:Rfns}}
\end{table*}
The last expression of Eq.~(\ref{eq:2-21}) is the formula which is
practically useful for numerical evaluations. The other expressions
are shown to indicate origins of the integrals.

If the second-order bias function $c^{(2)}_X(\bm{k}_1,\bm{k}_2)$ only
depends on magnitudes of wavevectors $k_1$ and $k_2$, and not on the
relative angle $\mu_{12}=\hat{\bm{k}_1}\cdot\hat{\bm{k}_2}$, the
fourth function generically vanishes: $R^X_4(k)=0$. If the first-order
bias function $c^{(1)}_X$ is scale-independent, it is explicitly shown
from the last expressions that $R^X_3(k) =[R_1(k) + R_2(k)]
c^{(1)}_X$. Specifically, the functions $R^X_3(k)$ and $R^X_4(k)$ are
redundant in the Lagrangian {\em local} bias models, in which
renormalized bias functions $c^{(n)}_X$ are scale-independent. This is
the reason only two functions $R_1(k)$ and $R_2(k)$ are needed in
Ref.~\cite{mat08b}. In general situations with Lagrangian {\em
  nonlocal} bias models, all four functions are needed. In a simple
model of halo bias in this paper, the second-order bias function
$c^{(2)}_X$ does not depend on the angle $\mu_{12}$ and $R_4(k)=0$ in
this case.

The second set of formulas is related to the convolution integrals of
the three-point propagators $\varGamma^{(2)}_X$ in calculating
the one-loop power spectrum. The integrals of the 
form,
\begin{align}
  \int_{\bm{k}_{12}=\bm{k}}
  {\cal F}(\bm{k}_1,\bm{k}_2) P_{\rm L}(k_1) P_{\rm L}(k_2),
\label{eq:2-21-1}
\end{align}
where ${\cal F}$ consists of LPT kernels $\bm{L}_n$ and
renormalized bias functions $c^{(n)}_X$ and $c^{(n)}_Y$, are reduced
to two-dimensional integrals, $Q^{XY}_n(k)$. The explicit formulas are
given Table~\ref{tab:Qformulas}.
\begin{table*}
\begin{tabular}{c|c|c}
    $\displaystyle {\cal F}(\bm{k}_1,\bm{k}_2)$
    & $\displaystyle\vphantom{\frac{\int}{\int}}
    \int_{\bm{k}_{12}=\bm{k}}
    {\cal F}(\bm{k}_1,\bm{k}_2) P_{\rm L}(k_1) P_{\rm L}(k_2)$
    & Diagram
    \\[.5pc] \hline\hline
    $\displaystyle L^{(2)}_i(\bm{k}_1,\bm{k}_2) L^{(2)}_j(\bm{k}_1,\bm{k}_2) $
    & $\displaystyle \vphantom{\frac{\int}{\int}}
    \frac{9}{49} \frac{k_ik_j}{k^4} Q_1(k) $
    &
    \includegraphics[height=1.8pc]{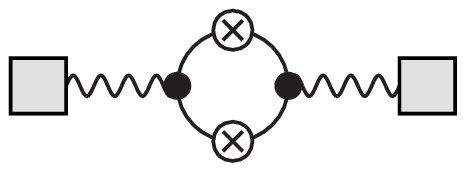}
    \\ [.5pc]\hline
    $\displaystyle L^{(1)}_i(\bm{k}_1) L^{(1)}_j(\bm{k}_2)
    L^{(2)}_k(\bm{k}_1,\bm{k}_2) $
    & $\displaystyle\vphantom{\frac{\int}{\int}}
    \frac{3}{14} \frac{(k_ik_j - k^2 \delta_{ij})k_k}{k^6}
    Q_1(k) + \frac{3}{7} \frac{k_ik_jk_k}{k^6} Q_2(k) $
    &
    \includegraphics[height=1.8pc]{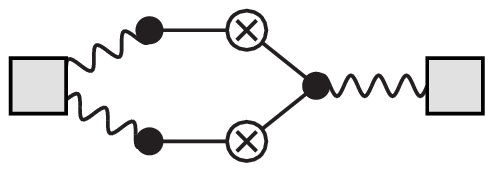}
    \\[.5pc]\hline
    $\displaystyle  L^{(1)}_{(i}(\bm{k}_1) L^{(1)}_j(\bm{k}_1)
    L^{(1)}_k(\bm{k}_2) L^{(1)}_{l)}(\bm{k}_2) $
    & $\displaystyle
    \begin{array}{l}\displaystyle\vphantom{\frac{\int}{\int}}
        \frac{3}{8}\,
        \frac{k_i k_j k_k k_l - 2k^2\delta_{(ij}k_k k_{l)} + k^4
          \delta_{(ij}\delta_{kl)}}{k^8} Q_1(k) \\
        \displaystyle\vphantom{\frac{\int}{\int}} \qquad
        - \frac12\,\frac{k_i k_j k_k k_l - k^2 \delta_{(ij}k_k
          k_{l)}}{k^8} Q_3(k)
        + \frac{k_i k_j k_k k_l}{k^8} Q_4(k) \end{array} $
    &
    \includegraphics[height=1.8pc]{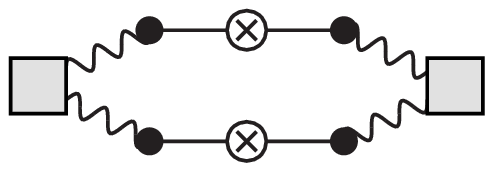}
    \\[.5pc]\hline
    $\displaystyle\vphantom{\frac{\int}{\int}}
    L^{(1)}_i(\bm{k}_1) L^{(2)}_j(\bm{k}_1,\bm{k}_2) c^{(1)}_X(k_2) $
    & $\displaystyle \frac{3}{7}\, \frac{k_ik_j}{k^4} Q^X_5(k) $
    &
    \includegraphics[height=1.8pc]{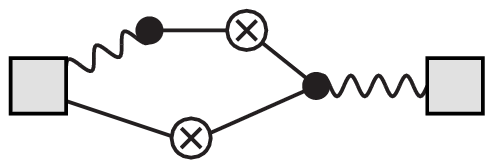}
    \\[.5pc]\hline
    $\displaystyle\vphantom{\frac{\int}{\int}}
    L^{(1)}_{(i}(\bm{k}_1) L^{(1)}_j(\bm{k}_1) L^{(1)}_{k)}(\bm{k}_2)
    c^{(1)}_X(k_2) $
    & $\displaystyle -\frac{1}{2}\,
    \frac{k_i k_j k_l - k^2 \delta_{(ij}k_{k)}}{k^6} Q^X_6(k) 
    + \frac{k_i k_j k_k}{k^6} Q^X_7(k) $
    &
    \includegraphics[height=1.8pc]{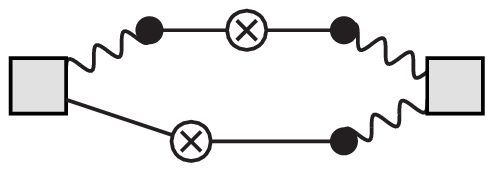}
    \\[.5pc]\hline
    $\displaystyle\vphantom{\frac{\int}{\int}}
    \bm{L}^{(2)}(\bm{k}_1,\bm{k}_2) c^{(2)}_X(\bm{k}_1,\bm{k}_2) $
    & $\displaystyle \frac37\frac{\bm{k}}{k^2} Q^X_8(k) $
    &
    \includegraphics[height=1.8pc]{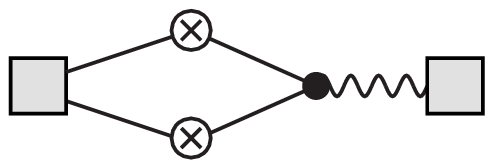}
    \\[.5pc]\hline
    $\displaystyle\vphantom{\frac{\int}{\int}}
    L^{(1)}_i(\bm{k}_1) L^{(1)}_j(\bm{k}_2) c^{(2)}_X(\bm{k}_1,\bm{k}_2) $
    & $\displaystyle \frac12\,\frac{k_i k_j  - k^2 \delta_{ij}}{k^4} Q^X_8(k)
    + \frac{k_i k_j}{k^4} Q^X_9(k) $
    &
    \includegraphics[height=1.8pc]{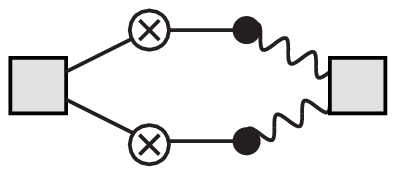}
    \\[.5pc]\hline
    $\displaystyle\vphantom{\frac{\int}{\int}}
    L^{(1)}_i(\bm{k}_1) L^{(1)}_j(\bm{k}_2) c^{(1)}_X(k_1) c^{(1)}_Y(k_2) $
    & $\displaystyle  \frac12\,\frac{k_i k_j  - k^2 \delta_{ij}}{k^4}
    Q^{XY}_{10}(k) + \frac{k_i k_j}{k^4} Q^{XY}_{11}(k) $
    &
    \includegraphics[height=1.8pc]{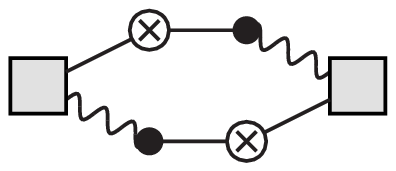}
    \\[.5pc]\hline
    $\displaystyle\vphantom{\frac{\int}{\int}}
    L^{(1)}_i(\bm{k}_1) L^{(1)}_j(\bm{k}_1) c^{(1)}_X(k_2) c^{(1)}_Y(k_2) $
    & $\displaystyle
    -\frac12\,\frac{k_i k_j  - k^2 \delta_{ij}}{k^4} Q^{XY}_{12}(k)
    + \frac{k_i k_j}{k^4} Q^{XY}_{13}(k) $
    &
    \includegraphics[height=1.8pc]{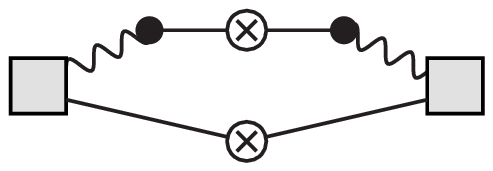}
    \\[.5pc]\hline
    $\displaystyle\vphantom{\frac{\int}{\int}}
    \bm{L}^{(1)}(\bm{k}_1) c^{(1)}_X(k_2)
    c^{(2)}_Y(\bm{k}_1,\bm{k}_2) $
    & $\displaystyle  \frac{\bm{k}}{k^2} Q^{XY}_{14}(k) $
    &
    \includegraphics[height=1.8pc]{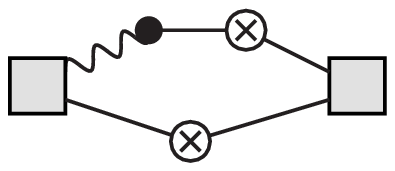}
    \\[.5pc]\hline
    $\displaystyle\vphantom{\frac{\int}{\int}}
    c^{(2)}_X(\bm{k}_1, \bm{k}_2) c^{(2)}_Y(\bm{k}_1, \bm{k}_2) $
    & $\displaystyle Q^{XY}_{15}(k) $
    &
    \includegraphics[height=1.8pc]{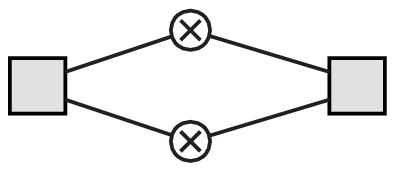}
\end{tabular}
\caption{Integral formulas for one-loop corrections, which are related
  to convolving three-point propagators.  In the  third and fifth
  formulas, the spatial indices are completely symmetrized. We denote
  $Q_n(k) = Q^{XY}_n(k)$ for $n=1,2,3,4$, as these functions are
  independent on the bias, and
  $Q^X_n(k) = Q^{XY}_n(k)$ for $n=5,6,7,8,9$, as these functions are
  only dependent on the bias of objects $X$.
  \label{tab:Qformulas}}
\end{table*}
For the third and fifth formulas in this Table, the indices of the LPT
kernels are symmetrized, since only symmetric combinations are used in
this paper. In this Table, we denote $Q_n(k) = Q^{XY}_n(k)$ for
$n=1,2,3,4$, as these functions are independent on the bias, and
$Q^X_n(k) = Q^{XY}_n(k)$ for $n=5,6,7,8,9$, as these functions are
only dependent on the bias of objects $X$.

The functions $Q^{XY}_n(k)$ are defined by two equivalent sets of
equations,
\begin{align}
  Q^{XY}_n(k) &= \int_{\bm{k}_{12} = \bm{k}}
  {\cal Q}^{XY}_n(\bm{k}_1,\bm{k}_2) P_{\rm L}(k_1)P_{\rm L}(k_2)
\nonumber\\
  &= 
  \frac{k^3}{4\pi^2}
  \int_0^\infty dr \int_{-1}^1 dx\,
  \tilde{\cal Q}^{XY}_n(r,x) P_{\rm L}(kr)
\nonumber\\
  &\hspace{7pc}\times
  P_{\rm L}\left(k\sqrt{1+r^2-2rx}\right),
\label{eq:2-22}
\end{align}
where integrands ${\cal Q}^{XY}_n(\bm{k}_1,\bm{k}_2)$, $\tilde{\cal
  Q}^{XY}_n(r,x)$ are given in Table~\ref{tab:Qfns}.
\begin{table}
\begin{tabular}{c|c|c}
    $n$
    & $\begin{array}{c} {\cal Q}^{XY}_n(\bm{k}_1,\bm{k}_2) \\
        \left[\bm{k}=\bm{k}_1+\bm{k}_2\right] \end{array}$
    & $\begin{array}{c}
        \tilde{\cal Q}^{XY}_n(r,x)\\
        \left[\begin{array}{c}
                y=(1+r^2-2rx)^{1/2}, \\ \mu=(x-r)/y \end{array} 
        \right] \end{array}  $
    \\[.3pc] \hline
    $1$
    & $\displaystyle\vphantom{\frac{\int}{\int}}
    \left[1 -
        \left(\frac{\bm{k}_1\cdot\bm{k}_2}{k_1k_2}\right)^2\right]^2 $
    & $\displaystyle \frac{r^2(1-x^2)^2}{y^4} $
    \\[.7pc]
    $2$
    & $\displaystyle\vphantom{\frac{\int}{\int}}
    \frac{(\bm{k}\cdot\bm{k}_1)(\bm{k}\cdot\bm{k}_2)}{{k_1}^2{k_2}^2}
    \left[1 - \left(\frac{\bm{k}_1\cdot\bm{k}_2}{k_1k_2}\right)^2\right] $
    & $\displaystyle  \frac{rx(1-rx)(1-x^2)}{y^4} $
    \\[.7pc]
    $3$
    & $\displaystyle\vphantom{\frac{\int}{\int}}
    \frac{k^4-6(\bm{k}\cdot\bm{k}_1)(\bm{k}\cdot\bm{k}_2)}{{k_1}^2{k_2}^2}
    \left[1 - \left(\frac{\bm{k}_1\cdot\bm{k}_2}{k_1k_2}\right)^2\right] $
    & $\displaystyle  \frac{(1-6rx+6r^2x^2)(1-x^2)}{y^4} $
    \\[.7pc]
    $4$
    & $\displaystyle\vphantom{\frac{\int}{\int}}
    \frac{(\bm{k}\cdot\bm{k}_1)^2(\bm{k}\cdot\bm{k}_2)^2}{{k_1}^4{k_2}^4}  $
    & $\displaystyle  \frac{x^2(1-rx)^2}{y^4} $
    \\[.7pc]
    $5$
    & $\displaystyle\vphantom{\frac{\int}{\int}}
    \frac{\bm{k}\cdot\bm{k}_1}{{k_1}^2}
    \left[1 - \left(\frac{\bm{k}_1\cdot\bm{k}_2}{k_1k_2}\right)^2\right]
    c^{(1)}_X(k_2) $
    & $\displaystyle \frac{rx(1-x^2)}{y^2} c^{(1)}_X(ky) $ 
    \\[.7pc]
    $6$
    & $\displaystyle\vphantom{\frac{\int}{\int}}
    \frac{k^2 - 3\bm{k}\cdot\bm{k}_1}{{k_1}^2}
    \left[1 - \left(\frac{\bm{k}_1\cdot\bm{k}_2}{k_1k_2}\right)^2\right]
    c^{(1)}_X(k_2) $
    & $\displaystyle \frac{(1-3rx)(1-x^2)}{y^2} c^{(1)}_X(ky) $
    \\[.7pc]
    $7$
    & $\displaystyle\vphantom{\frac{\int}{\int}}
    \left(\frac{\bm{k}\cdot\bm{k}_1}{{k_1}^2}\right)^2
    \frac{\bm{k}\cdot\bm{k}_2}{{k_2}^2}
    c^{(1)}_X(k_2) $
    & $\displaystyle \frac{x^2(1-rx)}{y^2} c^{(1)}_X(ky) $
    \\[.7pc]
    $8$
    & $\displaystyle\vphantom{\frac{\int}{\int}}
    \left[1 - \left(\frac{\bm{k}_1\cdot\bm{k}_2}{k_1k_2}\right)^2\right]
    c^{(2)}_X(\bm{k}_1,\bm{k}_2) $ 
    & $\displaystyle \frac{r^2(1-x^2)}{y^2} c^{(2)}_X(kr,ky;\mu) $
    \\[.7pc]
    $9$
    & $\displaystyle\vphantom{\frac{\int}{\int}}
    \frac{(\bm{k}\cdot\bm{k}_1)(\bm{k}\cdot\bm{k}_2)}{{k_1}^2{k_2}^2}
    c^{(2)}_X(\bm{k}_1,\bm{k}_2) $ 
    & $\displaystyle \frac{rx(1-rx)}{y^2} c^{(2)}_X(kr,ky;\mu) $
    \\[.7pc]
    $10$
    & $\displaystyle\vphantom{\frac{\int}{\int}}
    \left[1 - \left(\frac{\bm{k}_1\cdot\bm{k}_2}{k_1k_2}\right)^2\right]
    c^{(1)}_X(k_1) c^{(1)}_Y(k_2) $
    & $\displaystyle \frac{r^2(1-x^2)}{y^2} c^{(1)}_X(kr) c^{(1)}_Y(ky) $
    \\[.7pc]
    $11$
    & $\displaystyle\vphantom{\frac{\int}{\int}}
    \frac{(\bm{k}\cdot\bm{k}_1)(\bm{k}\cdot\bm{k}_2)}{{k_1}^2{k_2}^2}
    c^{(1)}_X(k_1) c^{(1)}_Y(k_2) $
    & $\displaystyle  \frac{rx(1-rx)}{y^2} c^{(1)}_X(kr) c^{(1)}_Y(ky) $
    \\[.7pc]
    $12$
    & $\displaystyle\vphantom{\frac{\int}{\int}}
    \frac{k^2}{{k_1}^2}
    \left[1 - \left(\frac{\bm{k}\cdot\bm{k}_1}{kk_1}\right)^2\right]
    c^{(1)}_X(k_2) c^{(1)}_Y(k_2) $
    & $\displaystyle \left(1 - x^2\right)
    c^{(1)}_X(ky)c^{(1)}_Y(ky) $
    \\[.7pc]
    $13$
    & $\displaystyle\vphantom{\frac{\int}{\int}}
    \left(\frac{\bm{k}\cdot\bm{k}_1}{{k_1}^2}\right)^2
    c^{(1)}_X(k_2)c^{(1)}_Y(k_2) $
    & $\displaystyle x^2 c^{(1)}_X(ky)c^{(1)}_Y(ky) $
    \\[.7pc]
    $14$
    & $\displaystyle\vphantom{\frac{\int}{\int}}
    \frac{\bm{k}\cdot\bm{k}_1}{{k_1}^2}
    c^{(1)}_X(k_2) c^{(2)}_Y(\bm{k}_1,\bm{k}_2) $
    & $\displaystyle rx\, c^{(1)}_X(ky) c^{(2)}_Y(kr,ky;\mu) $
    \\[.7pc]
    $15$
    & $\displaystyle\vphantom{\frac{\int}{\int}}
    c^{(2)}_X(\bm{k}_1,\bm{k}_2)c^{(2)}_Y(\bm{k}_1,\bm{k}_2) $
    & $\begin{array}{c}
        \displaystyle r^2\, c^{(2)}_X(kr,ky;\mu) \\
        \quad\qquad\times\,c^{(2)}_Y(kr,ky;\mu)
        \end{array}$
    \\[.7pc]
\end{tabular}
\caption{Integrands for functions $Q^{XY}_n(k)$ of Eq.~(\ref{eq:2-22}).
\label{tab:Qfns}}
\end{table}
The last expression of Eq.~(\ref{eq:2-22}) is the formula which is
practically useful for numerical evaluations. The first expressions
are shown to indicate origins of the integrands.

The third set of formulas is related to the initial bispectrum, which
is an indicator of primordial non-Gaussianity. The integrals of the
form,
\begin{align}
  \int_{\bm{k}_{12} = \bm{k}}
  {\cal F}(\bm{k}_1,\bm{k}_2) B_{\rm L}(k,k_1,k_2),
\label{eq:2-23}
\end{align}
where ${\cal F}$ consists of LPT kernels $\bm{L}_n$ and renormalized
bias functions $c^{(n)}_X$, are reduced to two-dimensional integrals,
$S^X_n(k)$. The explicit formulas are given in
Table~\ref{tab:Sformulas}.
\begin{table}
\begin{tabular}{c|c|c}
    $\displaystyle {\cal F}(\bm{k}_1,\bm{k}_2)$
    & $\displaystyle\vphantom{\frac{\int}{\int}}
    \int_{\bm{k}_{12}=\bm{k}}
    {\cal F}(\bm{k}_1,\bm{k}_2) B_{\rm L}(k,k_1,k_2)$
    & Diagram
    \\[.5pc] \hline\hline
    $\displaystyle\vphantom{\frac{\int}{\int}}
    \bm{L}^{(2)}(\bm{k}_1,\bm{k}_2) $
    & $\displaystyle\vphantom{\frac{\int}{\int}}
    \frac{3}{7} \frac{\bm{k}}{k^2} S_1(k) $
    &
    \includegraphics[height=1.5pc]{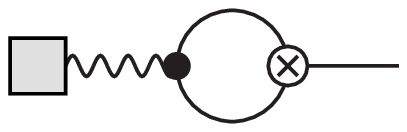}
    \\[.5pc] \hline
    $\displaystyle\vphantom{\frac{\int}{\int}}
    L_{1i}(\bm{k}_1) L_{1j}(\bm{k}_2) $
    & $\displaystyle\vphantom{\frac{\int}{\int}}
    \frac12\,\frac{k_ik_j-k^2\delta_{ij}}{k^4} S_1(k)
    + \frac{k_ik_j}{k^4} S_2(k) $
    &
    \includegraphics[height=1.5pc]{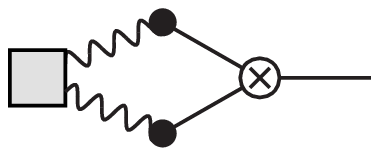}
    \\[.5pc] \hline
    $\displaystyle\vphantom{\frac{\int}{\int}}
    \bm{L}^{(1)}(\bm{k}_1) c^{(1)}_X(k_2) $
    & $\displaystyle \frac{\bm{k}}{k^2} S^X_3(k) $
    &
    \includegraphics[height=1.5pc]{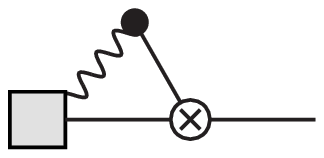}
    \\[.5pc] \hline
    $\displaystyle\vphantom{\frac{\int}{\int}}
    c^{(2)}_X(\bm{k}_1,\bm{k}_2) $
    & $\displaystyle S^X_4(k) $
    &
    \includegraphics[height=1.5pc]{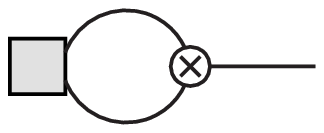}
\end{tabular}
\caption{Integral formulas for one-loop corrections, which are related
  to convolving three-point propagators with the linear bispectrum. We
  denote $S_1(k) = S^X_1(k)$ and  $S_2(k) = S^X_2(k)$, as these
  functions are independent on the bias.
  \label{tab:Sformulas}}
\end{table}
In this Table, we denote $S_1(k) = S^X_1(k)$ and $S_2(k) = S^X_2(k)$,
as these functions are independent on the bias. The functions
$S^X_n(k)$ are defined by two equivalent sets of equations,
\begin{align}
  S^X_n(k) &= \int_{\bm{k}_{12} = \bm{k}}
  {\cal S}^X_n(\bm{k}_1,\bm{k}_2) B_{\rm L}(k,k_1,k_2)
\nonumber\\
  &= 
  \frac{k^3}{4\pi^2}
  \int_0^\infty dr \int_{-1}^1 dx\,
  \tilde{\cal S}^X_n(r,x)
\nonumber\\
  &\hspace{7pc}\times
  B_{\rm L}\left(k,kr,k\sqrt{1+r^2-2rx}\right),
\label{eq:2-24}
\end{align}
where integrands ${\cal S}^X_n(\bm{k}_1,\bm{k}_2)$, $\tilde{\cal
  S}^X_n(r,x)$ are given in Table~\ref{tab:Sfns}.
\begin{table}
\begin{tabular}{c|c|c}
    $n$
    & $\begin{array}{c} {\cal S}^X_n(\bm{k}_1,\bm{k}_2) \\
        \left[\bm{k}=\bm{k}_1+\bm{k}_2\right] \end{array}$
    & $\begin{array}{c}
        \tilde{\cal S}^X_n(r,x)\\
        \left[\begin{array}{l}
                y=(1+r^2-2rx)^{1/2},\\ \mu=(x-r)/y \end{array} 
        \right] \end{array}  $
    \\[.3pc] \hline
    $1$
    & $\displaystyle\vphantom{\frac{\int}{\int}}
    1 - \left(\frac{\bm{k}_1\cdot\bm{k}_2}{k_1k_2}\right)^2 $
    & $\displaystyle \frac{r^2(1-x^2)}{y^2} $
    \\[.7pc]
    $2$
    & $\displaystyle\vphantom{\frac{\int}{\int}}
    \frac{(\bm{k}\cdot\bm{k}_1)(\bm{k}\cdot\bm{k}_2)}{{k_1}^2{k_2}^2} $
    & $\displaystyle \frac{rx(1-rx)}{y^2} $
    \\[.7pc]
    $3$
    & $\displaystyle\vphantom{\frac{\int}{\int}}
    \frac{\bm{k}\cdot\bm{k}_1}{{k_1}^2}
    c^{(1)}_X(k_2) $
    & $\displaystyle rx\, c^{(1)}_X(ky) $
    \\[.7pc]
    $4$
    & $\displaystyle\vphantom{\frac{\int}{\int}}
    c^{(2)}_X(\bm{k}_1,\bm{k}_2) $
    & $\displaystyle r^2\,c^{(2)}_X(kr,ky;\mu) $
    \\[.7pc]
\end{tabular}
\caption{Integrands for functions $S^X_n(k)$ of Eq.~(\ref{eq:2-24}).
\label{tab:Sfns}}
\end{table}

\subsection{The power spectra in redshift space
\label{subsec:RedshiftSpacePS}
}

As all the necessary integral formulas are derived in the previous
subsection, we are ready to write down the explicit formula of the
power spectrum in redshift space. The decomposition of
Eq.~(\ref{eq:2-1}) is applicable in redshift space, and it is
sufficient to give the explicit expressions for the functions
$\varPi(\bm{k})$, $R_{XY}(\bm{k})$, $Q_{XY}(\bm{k})$, $S_{XY}(\bm{k})$
in redshift space. These functions depends on not only the magnitude
$k$ but also the direction relative to the lines of sight.

We employ the distant-observer approximation for the redshift-space
distortions, and the lines of sight are fixed in the direction of the
third axis, $\hat{\bm{z}}$. Lagrangian kernels are replaced according
to Eq.~(\ref{eq:1-12}) in the formulas of propagators
Eqs.~(\ref{eq:1-8}) and (\ref{eq:1-13}). In those formulas, the
Lagrangian kernels appear only in the form of
$\bm{k}\cdot\bm{L}^{(n)}$. With the linear mapping of
Eq.~(\ref{eq:1-12}), we have
\begin{equation}
  \bm{k}\cdot\bm{L}^{(n)} \rightarrow
  \bm{k}\cdot\bm{L}^{{\rm s}(n)} = 
  \left(\bm{k} + nf\mu k\hat{\bm{z}} \right)\cdot\bm{L}^{(n)},
\label{eq:2-101}
\end{equation}
where
\begin{equation}
 \mu = \hat{\bm{k}}\cdot\hat{\bm{z}},
\label{eq:2-102}
\end{equation}
and $\hat{\bm{k}}=\bm{k}/k$. Thus, in the distant-observer
approximation of this paper, the direction dependence comes into the
formulas only through the direction cosine of Eq.~(\ref{eq:2-102}). We
denote the functions of Eqs.~(\ref{eq:2-2a})--(\ref{eq:2-2c}) as
$R_{XY}(k,\mu)$, $Q_{XY}(k,\mu)$, $S_{XY}(k,\mu)$ in the following.

Substituting Eq.~(\ref{eq:2-101}) into Eqs.~(\ref{eq:1-7}),
(\ref{eq:1-8}) and (\ref{eq:1-13}), one can see that evaluations of
Eqs.~(\ref{eq:2-2a})--(\ref{eq:2-2c}) are straightforward by means of
the integral formulas in the previous subsection. The results are
explicitly presented in the following.

The vertex resummation function of Eq.~(\ref{eq:1-7}) can be evaluated
by applying the same technique of the previous section. The relevant
integral is
\begin{align}
&
  \int\frac{d^3p}{(2\pi)^3}
  \left[\bm{k}\cdot\bm{L}^{{\rm s}(1)}(\bm{p})\right]^2
  P_{\rm L}(p)
\nonumber\\
& \qquad
  = (k_i + f\mu k \hat{z}_i) (k_j + f\mu k \hat{z}_j)
  \int\frac{d^3p}{(2\pi)^3}
  \frac{p_i p_j}{p^4}
  P_{\rm L}(p),
\label{eq:2-102-1}
\end{align}
and the last integral is proportional to the Kronecker's delta. The
proportional factor is evaluated by taking contraction of the indices.
Consequently, we have
\begin{equation}
  \varPi(k,\mu) = 
  \exp\left\{-\left[1 + f(f+2)\mu^2\right]
      \frac{k^2}{12\pi^2} \int dp P_{\rm L}(p)
  \right\}.
\label{eq:2-103}
\end{equation}

The two-point propagator of Eq.~(\ref{eq:1-8}) with the substitution
of Eq.~(\ref{eq:2-101}) is evaluated by means of
Table~\ref{tab:Rformulas}, where $R_n(k)$ functions are defined by
Eq.~(\ref{eq:2-21}) and Table~\ref{tab:Rfns}. The result is given by
\begin{align}
  \hat{\varGamma}^{(1)}_X(k,\mu) &=
  1 + c^{(1)}_X + 
   \frac{5}{21} R_1 + \frac37 R_2
  + \frac37 R^X_3 - R^X_4
\nonumber\\
  & \quad + \,
  \left[
    1 + \frac57 R_1 + \frac97 R_2 + \frac67 R^X_3 - R^X_4
  \right] f \mu^2
\nonumber\\
  & \quad
    -\,\frac37 R_1 f^2 \mu^2
  +  \left[
    \frac37 R_1 + \frac67 R_2
  \right] f^2 \mu^4.
\label{eq:2-104}
\end{align}
The quantities $c^{(1)}_X$, $R_n$, $R^X_n$ on LHS are functions of
$k$, although the arguments are omitted. The component $R_{XY}$ of
Eq.~(\ref{eq:2-2a}) is straightforwardly obtained by the above result
of the two-point propagator:
\begin{equation}
  R_{XY}(k,\mu) =
  \hat{\varGamma}^{(1)}_X(k,\mu)
  \hat{\varGamma}^{(1)}_Y(k,\mu)
  P_{\rm L}(k).
\label{eq:2-105}
\end{equation}
The tree-level contribution of the above equation is given by
$(b_X+f\mu^2)(b_Y+f\mu^2)P_{\rm L}(k)$ where $b_X = 1 + c^{(1)}_X$,
and Kaiser's linear formula of redshift-space distortions for the
power spectrum \cite{kai87} is exactly reproduced. In calculating the
mass power spectrum, $X=Y={\rm m}$, we only need terms with $R_1(k)$
and $R_2(k)$ and other terms $R^X_3(k)$ and $R^X_4(k)$ vanish since
$c^{(n)}_X = 0$ for unbiased mass density field.

The component $Q_{XY}(k,\mu)$ of Eq.~(\ref{eq:2-2b}) is similarly
evaluated, while the number of terms are larger. The result is given
by
\begin{equation}
  Q_{XY}(k,\mu) = \frac12
  \sum_{n,m} \mu^{2n} f^m \left[q^{XY}_{nm}(k) + q^{YX}_{nm}(k)\right],
\label{eq:2-106}
\end{equation}
where%
\allowdisplaybreaks
\begin{align}
  q^{XY}_{00} &= 
  \frac{9}{98} Q_1 + \frac37 Q_2 + \frac12 Q_4
  + \frac67 Q^X_5
  + 2 Q^X_7 + \frac37 Q^X_8 + Q^X_9
\nonumber\\ & \quad
  + Q^{XY}_{11}
  + Q^{XY}_{13} + 2Q^{XY}_{14}
  + \frac12 Q^{XY}_{15},
\label{eq:2-107a}
\\
  q^{XY}_{11} &= 
  \frac{18}{49} Q_1 + \frac{12}{7} Q_2 + 2Q_4
  + \frac{18}{7} Q^X_5 + 6 Q^X_7 + \frac67 Q^X_8
\nonumber\\ & \quad
  + 2 Q^X_9 + 2 Q^{XY}_{11} + 2 Q^{XY}_{13} + 2Q^{XY}_{14},
\label{eq:2-107b}
\\
  q^{XY}_{12} &= 
  -\frac{3}{14} Q_1 + \frac14 Q_3
  + Q^X_6 - \frac12 Q^X_8
  - \frac12 Q^{XY}_{10} + \frac12 Q^{XY}_{12},
\label{eq:2-107c}
\\
  q^{XY}_{22} &= 
  \frac{57}{98} Q_1 + \frac{15}{7} Q_2 - \frac14 Q_3
  + 3 Q_4
  + \frac{12}{7} Q^X_5 - Q^X_6 
  + 6 Q^X_7
\nonumber\\ & \quad
 + \frac12 Q^X_8 + Q^X_9
  + \frac12 Q^{XY}_{10} + Q^{XY}_{11}
  - \frac12 Q^{XY}_{12} + Q^{XY}_{13},
\label{eq:2-107d}\\
  q^{XY}_{23} &= 
  - \frac37 Q_1 + \frac12 Q_3 + Q^X_6,
\label{eq:2-107e}\\
  q^{XY}_{24} &= 
  \frac{3}{16} Q_1,
\label{eq:2-107f}\\
  q^{XY}_{33} &= 
  \frac37 Q_1 + \frac67 Q_2 - \frac12 Q_3
  + 2 Q_4
  - Q^X_6 + 2 Q^X_7,
\label{eq:2-107g}\\
  q^{XY}_{34} &= 
  - \frac38 Q_1 + \frac14 Q_3,
\label{eq:2-107h}\\
  q^{XY}_{44} &= 
  \frac{3}{16} Q_1 - \frac14 Q_3 + \frac12 Q_4,
\label{eq:2-107i}
\end{align}
\allowdisplaybreaks[0]%
and other $q^{XY}_{nm}(k)$'s which are not listed above all vanish.
The quantities $Q_n$, $Q^X_n$, $Q^{XY}_n$ are functions of $k$,
although the arguments are omitted. The $Q_n$ functions of
$n=1,\ldots,4$, $10,\ldots,13,15$ are symmetric with respect to $X
\leftrightarrow Y$, while those of $n=5,\ldots,9,14$ are not. In
calculating cross power spectra, $X \ne Y$, the symmetrization with
respect to $XY$ in Eq.~(\ref{eq:2-106}) is necessary. In calculating
auto power spectra, $X = Y$, two terms in the square bracket in
Eq.~(\ref{eq:2-106}) are the same, and can be replaced by
$2q^{XX}_{nm}(k)$. In calculating the mass power spectrum, $X=Y={\rm
  m}$, we only need terms with $Q_1(k),\ldots,Q_4(k)$ and other terms
$Q^X_5(k),\ldots,Q^{XY}_{15}(k)$ all vanish since $c^{(n)}_X = 0$ for
unbiased mass density field.

The component $S_{XY}(k,\mu)$ of Eq.~(\ref{eq:2-2c}) is similarly
evaluated. The result is given by
\begin{multline}
  S_{XY}(k,\mu) = \frac12
  \hat{\varGamma}^{(1)}_X(k,\mu)
  \left[
    \frac37 S_1 + S_2 + 2 S^Y_3 + S^Y_4
  \right.
\\
  \left.
    + \left(
        \frac67 S_1 + 2 S_2 + 2 S^Y_3
      \right) f \mu^2
  \right.
\\
  \left.
    - \frac12 S_1 f^2 \mu^2
    + \left(
        \frac12 S_1 + S_2
      \right) f^2 \mu^4
  \right] + (X \leftrightarrow Y).
\label{eq:2-108}
\end{multline}
The quantities $S_n$, $S^X_n$ and $S^Y_n$ are functions of $k$,
although the arguments are omitted. The normalized two-point
propagator $\hat{\varGamma}^{(1)}_X(k,\mu)$ in Eq.~(\ref{eq:2-108})
can be replaced by the tree-level term, $1 + c^{(1)}_X + f\mu^2$,
because the rest of the factor is already of one-loop order.

All the necessary components to calculate the power spectrum of
Eq.~(\ref{eq:2-1}) in redshift space,
\begin{equation}
  P_{XY}(k,\mu) =
  \varPi^2(k,\mu)
  \left[
      R_{XY}(k,\mu) + Q_{XY}(k,\mu) + S_{XY}(k,\mu)
  \right],
\label{eq:2-109}
\end{equation}
are provided above, i.e., Eqs.~(\ref{eq:2-103}), (\ref{eq:2-105}),
(\ref{eq:2-106}) and (\ref{eq:2-108}). Numerical integrations of
Eqs.~(\ref{eq:2-21}), (\ref{eq:2-22}) and (\ref{eq:2-24}) are not
difficult, once the model of renormalized bias functions $c^{(n)}_X$
and primordial spectra $P_{\rm L}(k)$, $B_{\rm L}(k_1,k_2,k_3)$ are
given. The last term $S_{XY}(k)$ is absent in the case of Gaussian
initial conditions.

\subsection{Evaluating correlation functions}

We have derived full expressions of power spectra of biased tracers in
the one-loop approximation. The correlation functions are obtained by
Fourier transforming the power spectrum. In real space, the relation
between the correlation function $\xi_{XY}(r)$ and the power spectrum
$P_{XY}(k)$ is standard:
\begin{equation}
  \xi_{XY}(r) = \int_0^\infty \frac{k^2dk}{2\pi^2} j_0(kr) P_{XY}(k),
\label{eq:2-200}
\end{equation}
where $j_l(z)$ is the spherical Bessel function. For a numerical
evaluation, it is convenient to first tabulate the values of power
spectrum $P_{XY}(k)$ of Eq.~(\ref{eq:2-14}) in performing the
one-dimensional integration of Eq.~(\ref{eq:2-200}).

In redshift space, multipole expansions of the correlation function
are useful \cite{ham92,CFW94,ham98}. For reader's convenience, we
summarize here the set of equations which is useful to numerically
evaluate the correlation functions in redshift space from the iPT
formulas of power spectra derived above. The multipole expansion of
the power spectrum in redshift space, $P_{XY}(k,\mu)$, with respect to
the direction cosine relative to lines of sight has a form,
\begin{equation}
  P_{XY}(k,\mu) = \sum_{l=0}^\infty p_{XY}^l(k) P_l(\mu),
\label{eq:2-201}
\end{equation}
where $P_l(\mu)$ is the Legendre polynomial. Inverting the above
equation by the orthogonal relation of Legendre polynomials, the
coefficient $p_{XY}^l(k)$ is given by
\begin{equation}
  p_{XY}^l(k) = \frac{2l+1}{2} \int_{-1}^1 d\mu P_l(\mu) P_{XY}(k,\mu).
\label{eq:2-202}
\end{equation}
Because of the distant-observer approximation, the index $l$ only
takes even integers.

The dependence on the direction $\mu$ of our power spectrum,
$P_{XY}(k,\mu)$ of Eq.~(\ref{eq:2-109}), appears in forms of $\mu^{2n}
e^{-\alpha\mu^2}$ where $n=0,1,2,\ldots$ are non-negative integers. It
is possible to analytically reduce the integral of
Eq.~(\ref{eq:2-202}) by using an identity
\begin{equation}
  \int_{-1}^1 d\mu\,\mu^{2n} e^{-\alpha\mu^2}
  = \alpha^{-n-1/2} \gamma\left(n+\frac12,\alpha\right),
\label{eq:2-203}
\end{equation}
where $\gamma(z,p)$ is the lower incomplete gamma function defined by
\begin{equation}
  \gamma(z,p) = \int_0^p e^{-t} t^{z-1} dt.
\label{eq:2-204}
\end{equation}
Although the number of terms is large, it is straightforward to obtain
the analytic expression of $p_{XY}^l(k)$ of Eq.~(\ref{eq:2-202}) in
terms of $Q_n(k)$, $R_n(k)$, $S_n(k)$, $c_X^{(1)}(k)$, and the lower
incomplete gamma function. Computer algebra like {\sc Mathematica}
should be useful for that purpose. Alternatively, it is feasible to
numerically integrate the one-dimensional integral of
Eq.~(\ref{eq:2-202}) for each $k$, once the functions $Q_n(k)$,
$R_n(k)$, $S_n(k)$, $c_X^{(1)}(k)$ are precomputed and tabulated. The
latter method is much simpler than the former.

The multipole expansion of the correlation function in redshift space,
$\xi_{XY}(r,\mu)$, with respect to the direction cosine relative to
lines of sight is given by
\begin{align}
  \xi_{XY}(r,\mu) &= \sum_{l=0}^\infty \xi_{XY}^l(r) P_l(\mu),
\label{eq:2-205a}\\
  \xi_{XY}^l(r) &=
  \frac{2l+1}{2} \int_{-1}^1 d\mu P_l(\mu) \xi_{XY}(r,\mu).
\label{eq:2-205b}
\end{align}
Since the power spectrum $P_{XY}(k,\mu)$ and the correlation function
$\xi_{XY}(k,\mu)$ are related by a three-dimensional Fourier
transform, corresponding multipoles are related by \cite{ham98}
\begin{equation}
  \xi_{XY}^l(r) =
  i^{-l} \int_0^\infty \frac{k^2dk}{2\pi^2} j_l(kr) p_{XY}^l(k).
\label{eq:2-207}
\end{equation}
Since $l$ is an even integer, the above equation is a real number.
Once the multipoles of power spectrum $p_{XY}^l(k)$ are evaluated by
either method described above and tabulated as a function of $k$, we
have a multipoles of the correlation function $\xi_{XY}^l(r)$ by a
simple numerical integration of Eq.~(\ref{eq:2-207}). Because the
vertex resummation factor exponentially damps for high-$k$, the
numerical integration of Eq.~(\ref{eq:2-207}) is stable enough.

\subsection{A sample comparison with numerical simulations}

\begin{figure}[t]
\begin{center}
\includegraphics[width=19pc]{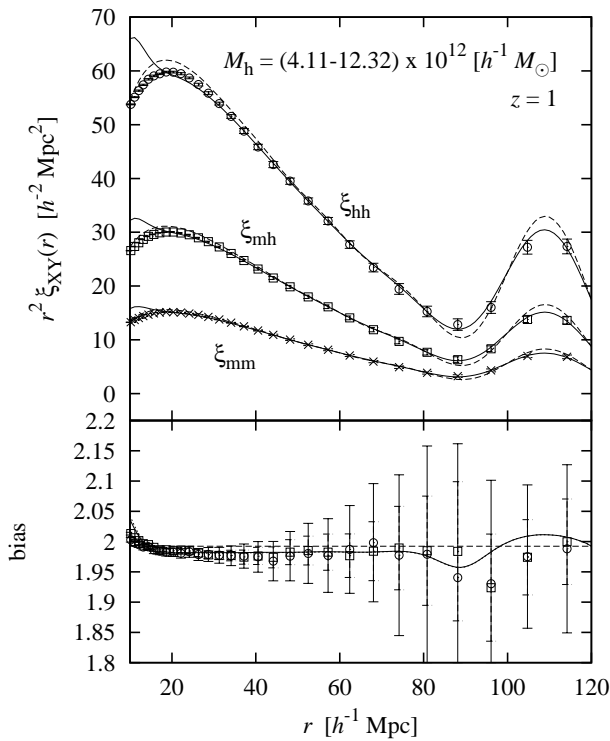}
\caption{\label{fig:XiHalo} The correlation functions in real space.
  The prediction of one-loop iPT is compared with numerical
  simulations. The results of mass auto-correlation,
  $\xi_\mathrm{mm}$, halo auto-correlation, $\xi_\mathrm{hh}$, and
  mass-halo cross-correlation, $\xi_\mathrm{mh}$ are compared in the
  above panel. Dashed lines represent the predictions of linear
  theory, solid lines represent those of one-loop iPT, and symbols
  with error bars represent the results of numerical simulations. In
  the bottom panel, scale-dependent bias parameters, which are defined
  by $\sqrt{\xi_\mathrm{hh}/\xi_\mathrm{mm}}$ for auto-correlations
  and $\xi_\mathrm{mh}/\xi_\mathrm{mm}$ for cross-correlations, are
  plotted. Predictions of iPT are given by solid line for
  auto-correlations and by dotted line for cross-correlations. These
  two lines are almost overlapped and indistinguishable. The
  horizontal dashed line corresponds to the prediction of linear
  theory with a constant bias factor.}
\end{center}
\end{figure}

The purpose of this paper is to analytically derive explicit formulas
of one-loop power spectra in iPT, and detailed analysis of numerical
consequences of derived formulas is beyond the scope of paper. In this
subsection, we only present a sample comparison with halos in $N$-body
simulations. In Fig.~\ref{fig:XiHalo}, correlation functions in real
space are presented.

The numerical halo catalogs in this figure are the same as the ones
used in Sato \& Matsubara (2011; 2013) \cite{SM11,SM13}. The
$N$-body simulations are performed by a publicly available
tree-particle mesh code, \textit{Gadget2} \cite{spr05} with
cosmological parameters $\varOmega_\mathrm{M} = 0.265$,
$\varOmega_\Lambda = 0.735$, $\varOmega_\mathrm{b} = 0.0448$, $h =
0.71$, $n_\mathrm{s} = 0.963$, $\sigma_8 = 0.80$. Other simulation
parameters are given by the box size $L_\mathrm{box} =
1000\,h^{-1}\mathrm{Mpc}$, the number of particles $N_\mathrm{p} =
1024^3$, initial redshift $z_\mathrm{ini} = 36$, the softening length
$r_\mathrm{s} = 50 h^{-1}\mathrm{kpc}$, and the number of realizations
$N_\mathrm{run} = 30$. Initial conditions are generated by a code
based on 2nd-order Lagrangian perturbation theory (2LPT) \cite{CPS06,
  VN11}, and initial spectrum is calculated by CAMB \cite{CAMB}. The
halos are selected by a Friends-of-Friends algorithm \cite{DEFW85}
with linking length of 0.2 times the mean separation. The output
redshift of the halo catalog is $z=1.0$, and the mass range of the
selected halos is $4.11 \times 10^{12} h^{-1}M_\odot \leq M \leq 12.32
\times 10^{12} h^{-1}M_\odot$.

In the upper panel, the auto- and cross-correlation functions of mass
and halos, $\xi_{\rm hh}$, $\xi_{\rm mh}$, $\xi_{\rm mm}$, are
plotted. Since the amplitude of linear halo bias, $b^\mathrm{L}_1$,
predicted by the peak-background split in the simple halo model, does
not accurately reproduce the value of halo bias in numerical
simulations, we consider the value of smoothing radius $R$ (or mass
$M$) in the simple model of the renormalized bias function as a free
parameter. We approximately treat this freely fitted radius as a
representative value, and ignore the finiteness of mass range, e.g.,
Eq.~(\ref{eq:1-40}). The same value of radius is used both in auto-
and cross-correlations, $\xi_\mathrm{hh}$ and $\xi_\mathrm{mh}$. We
use a Gaussian window function $W(kR) = e^{-k^2R^2/2}$, while the
shape of the window function does not change the predictions on large
scales. There is no fitting parameter for the mass auto-correlation
function $\xi_\mathrm{mm}$. As obviously seen in the Figure, the
predictions of one-loop iPT agree well with $N$-body simulations on
scales $\gtrsim 30\,h^{-1}\mathrm{Mpc}$ where the perturbation theory
is applicable.

In the lower panel, scale-dependent bias parameters are plotted. Two
definitions of linear bias factor,
$\sqrt{\xi_\mathrm{hh}/\xi_\mathrm{mm}}$ and
$\xi_\mathrm{mh}/\xi_\mathrm{mm}$, are presented. The iPT predicts
almost similar curves for both definitions, and slight
scale-dependence of linear bias on BAO scales is suggested. Such
scale-dependence is already predicted also in models of Lagrangian
local bias \cite{mat08b}. Unfortunately, the $N$-body simulations used
in this comparison are not sufficiently large to quantitatively
confirm the prediction for the scale-dependent bias. However, a recent
$N$-body analysis of the MICE Grand Challenge run \cite{Cro+13} shows
qualitatively the same scale-dependence. This observation exemplifies
unique potentials of the method of iPT.

\section{\label{sec:RelationPrevious}
Relation to previous work
}

\subsection{Lagrangian resummation theory}

It is worth mentioning here the relation between the above formulas
and previous results of Ref.~\cite{mat08b}, in which the Lagrangian
resummation theory (LRT) with local Lagrangian bias is developed. The
iPT is a superset of LRT. The results of Ref.~\cite{mat08b} can be
derived from the formulas in this paper by restricting to the local
Lagrangian bias and by neglecting contributions from the primordial
non-Gaussianity, although the way to derive the same results is
apparently different. The definitions of $Q_n$ and $R_n$ functions are
somehow different in Ref.~\cite{mat08b} from those in this paper. The
notational correspondences are summarized in Table~\ref{tab:PrevRes}.
\begin{table}
\begin{tabular}{c|c|c}
  This paper & Ref.~\cite{mat08b} & Ref.~\cite{mat12} \\ \hline
  $R_1(k)$ & $R_1(k)/P_{\rm L}(k)$ & - \\
  $R_2(k)$ & $R_2(k)/P_{\rm L}(k)$ & - \\[.2pc]
  $R_3^X(k)$ &
  $\langle F'\rangle \left[R_1(k)+R_2(k)\right]/P_{\rm L}(k)$ & -\\[.2pc]
  $R_4^X(k)$ & $0$ & - \\ \hline
  $Q_1(k)$ & $Q_1(k)$ & - \\
  $Q_2(k)$ & $Q_2(k)$ & - \\
  $Q_3(k)$ & $Q_4(k) - 6 Q_2(k)$ & - \\
  $Q_4(k)$ & $Q_3(k)$ & - \\[.2pc]
  $Q_6^X(k)$ & $\langle F'\rangle Q_6(k)$ & - \\[.2pc]
  $Q_7^X(k)$ & $\langle F'\rangle Q_7(k)$ & - \\[.2pc]
  $Q_8^X(k)$ & $\langle F''\rangle Q_8(k)$ & - \\[.2pc]
  $Q_9^X(k)$ & $\langle F''\rangle Q_9(k)$ & - \\[.2pc]
  $Q_{10}^{XX}(k)$ & $\langle F'\rangle^2 Q_8(k)$ & - \\[.2pc]
  $Q_{11}^{XX}(k)$ & $\langle F'\rangle^2 Q_9(k)$ & - \\[.2pc]
  $Q_{12}^{XX}(k)$ & $\langle F'\rangle^2 Q_{10}(k)$ & - \\[.2pc]
  $Q_{13}^{XX}(k)$ & $\langle F'\rangle^2 Q_{11}(k)$ & - \\[.2pc]
  $Q_{14}^{XX}(k)$ & $\langle F'\rangle\langle F''\rangle Q_{12}(k)$ & - \\[.2pc]
  $Q_{15}^{XX}(k)$ & $\langle F''\rangle^2 Q_{13}(k)$ & - \\ \hline
  $S_1(k)$ & - & $R_2(k)$ \\
  $S_2(k)$ & - & $2R_1(k) - R_2(k)$ \\[.2pc]
  $S_3^X(k)$ & - & $Q_1(k)/2$\\[.2pc]
  $S_4^X(k)$ & - & $Q_2(k)$ \\ \hline
\end{tabular}
\caption{
  When the local Lagrangian bias is employed, and
  primordial non-Gaussianity is not considered, the expression of the
  auto power spectrum ($X=Y$) in this paper reproduces the result of
  Ref.~\cite{mat08b}. When contributions
  from the primordial non-Gaussianity are extracted, the results of
  Ref.~\cite{mat12} are reproduced.
  　Correspondences of the functions defined in this
  paper and those defined in Refs.~\cite{mat08b,mat12} are provided
  in this Table.
  The renormalized bias functions are constants in local bias models,
  and denoted by $\langle F'\rangle = c^{(1)}_X$ and $\langle
  F''\rangle = c^{(2)}_X$ in Ref.~\cite{mat08b}.
  \label{tab:PrevRes}}
\end{table}

In Ref.~\cite{mat08b}, the linear density field $\delta_{\rm L}$ and
the biased density field in Lagrangian space $\delta^{\rm L}_X$ are
related by a local relation $\delta^{\rm L}_X(\bm{q}) = F(\delta_{\rm
  L}(\bm{q}))$ in Lagrangian configuration space. Fourier transforming
this relation, the renormalized bias functions of Eq.~(\ref{eq:1-3})
in models of local Lagrangian bias reduce to scale-independent
parameters,
\begin{equation}
    c^{(n)}_X = \left\langle F^{(n)} \right\rangle,
\label{eq:3-1}
\end{equation}
where $F^{(n)} = \partial^n F/\partial {\delta_{\rm L}}^n$ is the
$n$th derivative of the function $F(\delta_{\rm L})$. Thus the
renormalized bias functions are independent on wavevectors in the
case of local bias, and we have $\langle F'\rangle = c^{(1)}_X$ and
$\langle F''\rangle = c^{(2)}_X$, etc.

It is explicitly shown that the results of Ref.~\cite{mat08b} are
exactly reproduced by setting $X=Y$ and $S_{XY}=0$, expanding the
product $(\hat{\varGamma}^{(1)}_X)^2$ in $R_{XX}$ and adopting the
replacement of variables according to the Table~\ref{tab:PrevRes}. In
making such a comparison, the product
$\varGamma^{(1)}_X\varGamma^{(1)}_Y$ should be expanded up to the
second-order terms in $P_{\rm L}(k)$ (i.e., one-loop terms). Thus,
Eq.~(\ref{eq:2-1}) is considered as a nontrivial generalization of the
previous formula of Ref.~\cite{mat08b}. Another previous formula of
Ref.~\cite{mat08a} is a special case of Ref.~\cite{mat08b} without
biasing. As a consequence, setting $c^{(n)}_X=0$, $S_{XY} = 0$ in
Eq.~(\ref{eq:2-1}) reproduces the results of Ref.~\cite{mat08a}.

\subsection{Scale-dependent bias and primordial non-Gaussianity}

Contributions from the primordial bispectrum, if any, are included in
$S_{XY}$. In the cases of $X = Y$ and $X \ne Y = {\rm m}$, the
relations between the primordial bispectrum and scale-dependent bias
are already analyzed in Ref.~\cite{mat12} with generally nonlocal
Lagrangian bias. In the presence of primordial bispectrum, the
scale-dependent bias emerges on very large scales \cite{dal08,MV08}.
The iPT generalizes the previous formulas of the scale-dependent bias
with less number of approximations. The previous formulas of
scale-dependent bias \cite{dal08,SK10,DJS11a,DJS11b}, which are
derived in the approximation of peak-background split for the halo
bias, are exactly reproduced as limiting cases of the formula derived
by iPT \cite{mat12}. It should be noted that the formula of
scale-dependent bias in the framework of iPT is not restricted to the
particular model of halo bias. Therefore the iPT provides the most
general formula of the scale-dependent bias among previous work. The
correspondence between the functions defined in Ref.~\cite{mat12} and
those in this paper is summarized in Table~\ref{tab:PrevRes}.

In this paper, the cross power spectrum of two differently biased
objects, $X$ and $Y$ are considered in general. One can derive the
scale-dependent bias of cross power spectrum $P_{XY}(k)$ as
illustrated below. In the following argument, the redshift-space
distortions are neglected for simplicity, although it is
straightforward to include them. We define the scale-dependent bias
$\varDelta b_{XY}$ of cross power spectrum by
\begin{equation}
  P_{XY}(k) = \left[b_{XY}(k) + \varDelta b_{XY}(k)\right]^2
  P_{\rm m}(k),
\label{eq:3-10}
\end{equation}
where $P_{\rm m}(k)$ is the matter power spectrum, and $b_{XY}(k)$ is
the linear bias factor of the cross power spectrum without
contributions from primordial non-Gaussianity. In the lowest-order
approximation, $b_{XY}(k) = [b_X(k) b_Y(k)]^{1/2}$, where $b_X(k)$ and
$b_Y(k)$ are linear bias factors of objects $X$ and $Y$, respectively.
When higher orders of $\varDelta b_{XY}$ are neglected, we have
\begin{equation}
  \varDelta b_{XY} = \frac12 b_{XY}(k)
  \left[\frac{\varDelta P_{XY}(k)}{P^{\rm G}_{XY}(k)}
      - \frac{\varDelta P_{\rm m}(k)}{P^{\rm G}_{\rm m}(k)}
  \right],
\label{eq:3-11}
\end{equation}
where $P^{\rm G}_{XY}(k)$ and $P^{\rm G}_{\rm m}(k)$ are the Gaussian
parts of cross power spectrum and the auto power spectrum of mass,
respectively, and $\varDelta P_{XY}(k)$ and $\varDelta P_{\rm m}(k)$
are corresponding contributions from primordial non-Gaussianity so
that the full spectra are given by $P_{XY}(k) = P^{\rm G}_{XY}(k) +
\varDelta P_{XY}(k)$ and $P_{\rm m}(k) = P^{\rm G}_{\rm m}(k) +
\varDelta P_{\rm m}(k)$.

On sufficiently large scales, nonlinear gravitational evolutions are
not important, and dominant contributions to the multi-point
propagators are asymptotically given by \cite{mat12}
\begin{align}
  \hat{\varGamma}^{(1)}_X(k) &\approx
  b_X(k),
\label{eq:3-12a}\\
  \hat{\varGamma}^{(2)}_X(\bm{k}_1,\bm{k}_2) &\approx
  c^{(2)}_X(\bm{k}_1,\bm{k}_2),
\label{eq:3-12b}
\end{align}
where $b_X(k) = 1 + c^{(1)}_X(k)$ is the linear bias factor of object
$X$. In this limit, Eq.~(\ref{eq:2-2c}) reduces to
\begin{equation}
    S_{XY}(k) \approx
    b_X(k)
    \int_{\bm{k}_{12}=\bm{k}} c^{(2)}_Y(\bm{k}_1,\bm{k}_2)
    B_{\rm L}(k,k_1,k_2).
\label{eq:3-13}
\end{equation}
In the lowest-order approximation with a large-scale limit, the
predictions of iPT are given by
\begin{align}
  P^{\rm G}_{\rm m}(k) &\approx P_{\rm L}(k),
 & P^{\rm G}_{XY}(k) &\approx b_X(k)b_Y(k)P_{\rm L}(k),
\label{eq:3-14a}\\
  \varDelta P_{\rm m}(k) &\approx 0,
  & \varDelta P_{XY}(k) &\approx S_{XY}(k),
\label{eq:3-14b}
\end{align}
and we have $b_{XY}(k) = [b_X(k)b_Y(k)]^{1/2}$ as previously noted.
Substituting these equations into Eq.~(\ref{eq:3-11}), we have
\begin{equation}
  \varDelta b_{XY}(k) \approx
  \frac{S_{XY}(k)}{2\sqrt{b_X(k)b_Y(k)}\, P_{\rm L}(k)}.
\label{eq:3-15}
\end{equation}
This equation gives the general formula of the scale-dependent bias
for cross power spectra in general.

In a case of the auto-power spectrum with $X=Y$, the above equation
reduces to a known result \cite{mat12}, $\varDelta b_X \approx
S_{XX}(k)/[2b_X(k)P_{\rm L}(k)]$. Previous formulas of the
scale-dependent bias in the approximation of peak-background split are
reproduced in limiting cases of this result, adopting the renormalized
bias functions $c^{(n)}_X$ in the nonlocal model of halo bias
described in Sec.~\ref{subsec:HaloBias}. The integral of
Eq.~(\ref{eq:3-13}) is scale-dependent according to the squeezed limit
of the primordial bispectrum, $B_{\rm L}(k,k_1,k_2)$ with $k \ll
k_1,k_2$. Thus, the scale-dependencies of the bias in cross power
spectra are similar to those in auto power spectra. Amplitudes of the
scale-dependent bias are different. When the primordial
non-Gaussianity are actually detected, scale-dependent biases of cross
power spectra of multiple kinds of objects would be useful to
cross-check the detection.

\subsection{Convolution Lagrangian perturbation theory
\label{subsec:CLPT}
}

Recently, a further resummation method, called the convolution
Lagrangian perturbation theory (CLPT) \cite{CRW13}, is proposed on the
basis of LRT. The implementation of the CLPT actually improves the
nonlinear behavior on small scales where the original LRT breaks down.
The proposed CLPT is based on the LRT in which only local Lagrangian
bias can be incorporated.

\begin{figure}
\begin{center}
\includegraphics[width=12pc]{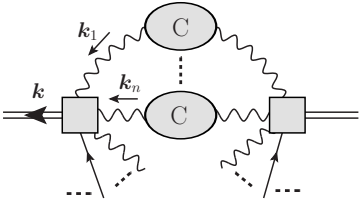}
\caption{\label{fig:CLPT} Diagrammatic representation of the
  resummation scheme of CLPT \cite{CRW13}. The original CLPT does not
  include the effects of nonlocal bias, and can be easily extended to
  include them by applying the formalism of iPT and the resummation of
  this type of diagrams.}
\end{center}
\end{figure}

\begin{figure}
\begin{center}
\includegraphics[width=20pc]{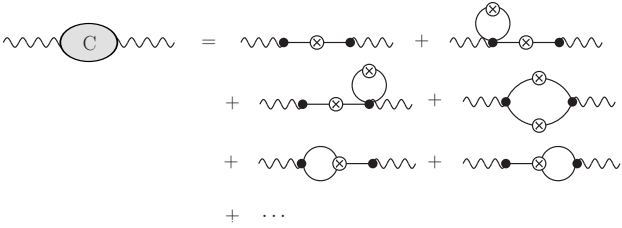}
\caption{\label{fig:Cpiece} Ingredients of the displacement
  correlator. All the diagrams up to one-loop approximation are
  shown. These diagrams are resummed in CLPT.}
\end{center}
\end{figure}

Under the light of iPT, the resummation scheme of CLPT corresponds to
resumming the diagrams depicted by Fig.~\ref{fig:CLPT}.
The shaded ellipse with the symbol `C' represents a summation of all
the possible connected diagrams. The actual ingredients are shown in
Fig.~\ref{fig:Cpiece} up to the one-loop approximation.
The corresponding function of this Figure is given by
\begin{multline} \tilde{\varLambda}_{ij}(\bm{k}) = - L^{(1)}_i(\bm{k})
    L^{(1)}_j(\bm{k}) P_{\rm L}(k)
    \\
    - \int \frac{d^3p}{(2\pi)^3} L^{(1)}_{(i}(\bm{k})
    L^{(3)}_{j)}(\bm{k},\bm{p},-\bm{p}) P_{\rm L}(p) P_{\rm L}(k)
    \\
    - \frac12 \int_{\bm{k}_{12}=\bm{k}} L^{(2)}_i(\bm{k}_1,\bm{k}_2)
    L^{(2)}_j(\bm{k}_1,\bm{k}_2) P_{\rm L}(k_1) P_{\rm L}(k_2)
    \\
    - L^{(1)}_{(i}(\bm{k}) \int_{\bm{k}_{12}=\bm{k}}
    L^{(2)}_{j)}(\bm{k}_1,\bm{k}_2) B_{\rm L}(k,k_1,k_2).
\label{eq:3-20}
\end{multline}
The indices $i,j$ are symmetrized on RHS of the above equation. This
function is the same as $C_{ij}(\bm{k})$ in Ref.~\cite{mat08b}, and
$-C_{ij}(\bm{k})$ in Ref.~\cite{mat08a}. We refer to the graph of
Fig.~\ref{fig:Cpiece} and Eq.~(\ref{eq:3-20}) as ``displacement
correlator'' below. To the full orders, the displacement correlator
$\tilde{\varLambda}_{ij}(\bm{k})$ is given by
\begin{equation}
  \left\langle
      \tilde{\varPsi}_i(\bm{k}) \tilde{\varPsi}_j(\bm{k}')
  \right\rangle_{\rm c}
  = - (2\pi)^3 \delta_{\rm D}^3(\bm{k}+\bm{k}')
  \tilde{\varLambda}_{ij}(\bm{k}).
\label{eq:3-21}
\end{equation}
The expression of Eq.~(\ref{eq:3-20}) is also obtained from this
equation, adopting the one-loop approximation in the perturbative
expansion of Eq.~(\ref{eq:1-9}).

Using the displacement correlator, the diagrams of Fig.~\ref{fig:CLPT}
can be represented by a convolution integral of the form
\begin{multline}
  \sum_{n=0}^\infty \frac{(-1)^n}{n!}
  k_{i_1} \cdots k_{i_n} k_{j_1} \cdots k_{j_n}
  \int_{\bm{k}_{1\cdots n}=\bm{k}'}
  \tilde{\varLambda}_{i_1 j_1}(\bm{k}_1) \cdots
  \tilde{\varLambda}_{i_n j_n}(\bm{k}_n)
\\
  = \int d^3q\,e^{-i\bm{k}_{1\cdots n}\cdot\bm{q}}
  \exp\left[-k_ik_j\varLambda_{ij}(\bm{q})\right],
\label{eq:3-22}
\end{multline}
where $\bm{k}$ is the wave vector of the nonlinear power spectrum
$P_{XY}(\bm{k})$ to evaluate, $\bm{k}_{1\cdots n} = \bm{k}_1 + \cdots
\bm{k}_n$ is the total wave vector that flows through the resummed
part of Fig.~\ref{fig:CLPT}, and
\begin{equation}
  \varLambda_{ij}(\bm{q}) = 
  \int \frac{d^3k}{(2\pi)^3} e^{i\bm{k}\cdot\bm{q}}
  \tilde{\varLambda}_{ij}(\bm{k})
\label{eq:3-23}
\end{equation}
is the displacement correlator in configuration space. The convolution
integral of Eq.~(\ref{eq:3-22}) contributes multiplicatively to the
evaluation of the power spectrum $P_{XY}(\bm{k})$. 

The displacement correlator in configuration space,
Eq.~(\ref{eq:3-23}), is given by the full-order displacement field
$\bm{\varPsi}(\bm{q})$ as
\begin{equation}
  \varLambda_{ij}(\bm{q}) = -
  \left\langle
      \varPsi_i(\bm{q}_2) \varPsi_j(\bm{q}_1)
  \right\rangle_{\rm c},
\label{eq:3-24}
\end{equation}
where $\bm{q} = \bm{q}_2 - \bm{q}_1$. This function is denoted as
$C_{ij}(\bm{q})/2$ in Ref.~\cite{CRW13}, and thus we have a
correspondence,
\begin{equation}
  C^{\rm CLPT}_{ij}(\bm{q}) = 2\varLambda_{ij}(\bm{q})
\label{eq:3-25}
\end{equation}
In the CLPT, the vertex resummation factor is included in a function
$A_{ij}(\bm{q}) = B_{ij} + C_{ij}(\bm{q})$ of their notation, where
$B_{ij} = 2\sigma_\eta^2\delta_{ij}$ and $\sigma_\eta^2 =
\langle|\bm{\varPsi}|^2\rangle/3$. Thus we have a correspondence, 
\begin{equation}
  A^{\rm CLPT}_{ij}(\bm{q})
  = \frac23 \sigma_\eta^2 \delta_{ij} + 2 \varLambda_{ij}(\bm{q}).
\label{eq:3-26}
\end{equation}
The first term in the LHS corresponds to the vertex resummation in
iPT, and is kept exponentiated in both original LRT and CLPT. The
second term is kept exponentiated in CLPT and expanded in the original
LRT formalism.

When the Lagrangian local bias is assumed ($c^{(1)}_X = \langle
F'\rangle$, $c^{(2)}_X = \langle F''\rangle$,...), and the convolution
resummation of Fig.~\ref{fig:CLPT} is taken into account in the iPT,
the formalism of CLPT is exactly reproduced. When the Lagrangian
nonlocal bias is allowed in the iPT with the convolution resummation,
we obtain a natural extension of the CLPT without restricting to
models of local Lagrangian bias.

\begin{figure}
\begin{center}
\includegraphics[width=20pc]{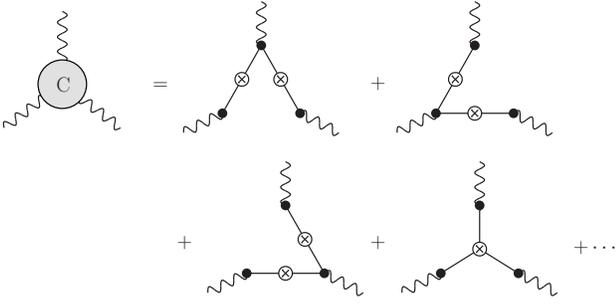}
\caption{\label{fig:Cthree} Connected diagrams with three wavy lines
  up to tree-level approximation. These diagrams are not resummed in
  CLPT.}
\end{center}
\end{figure}

Extending this diagrammatic understanding of CLPT in the framework of
iPT, it is possible to consider further convolution resummations that
are not included in the formulation of CLPT. In the CLPT, only
connected diagrams with two wavy lines (i.e., Fig.~\ref{fig:Cpiece})
are resummed. We define the three-point correlator of displacement,
$\tilde{\varLambda}_{ijk}(\bm{k}_1,\bm{k}_2,\bm{k}_3)$ where
$\bm{k}_1+\bm{k}_2+\bm{k}_3=\bm{0}$ by the connected diagrams with
three wavy lines as shown in Fig.~\ref{fig:Cthree}.
This function is given by
\begin{multline}
  \left\langle
      \tilde{\varPsi}_i(\bm{k}_1) \tilde{\varPsi}_j(\bm{k}_2)
      \tilde{\varPsi}_j(\bm{k}_3)
  \right\rangle_{\rm c}
\\
  = -i\,(2\pi)^3 \delta_{\rm D}^3(\bm{k}_1+\bm{k}_2+\bm{k}_3)
  \tilde{\varLambda}_{ijk}(\bm{k}_1,\bm{k}_2,\bm{k}_3).
\label{eq:3-27}
\end{multline}
to the full order. This three-point correlator
$\tilde{\varLambda}_{ijk}$ is the same as $-C_{ijk}$ in
Ref.~\cite{mat08b} and $-iC_{ijk}$ in Ref.~\cite{mat08a}. In a similar
way as Fig.~\ref{fig:CLPT} and Eq.~(\ref{eq:3-22}), including
resummations of the three-point correlator modifies the convolution
integral of Eq.~(\ref{eq:3-22}) as
\begin{equation}
  \int d^3q\,e^{-i\bm{k}'\cdot\bm{q}}
  \exp\left[
      -k_ik_j\varLambda_{ij}(\bm{q})
      + k_i k_j k_k \varLambda_{ijk}(\bm{q})
  \right],
\label{eq:3-28}
\end{equation}
where $\bm{k}'$ is the total wave vector that flows through the
resummed part, and
\begin{equation}
  \varLambda_{ijk}(\bm{q}) = 
  \int \frac{d^3k}{(2\pi)^3} e^{i\bm{k}\cdot\bm{q}}
  \int \frac{d^3p}{(2\pi)^3}
  \tilde{\varLambda}_{ijk}(\bm{k},-\bm{p},\bm{p}-\bm{k}).
\label{eq:3-29}
\end{equation}
One can similarly consider four- and higher-point convolution
resummations, which naturally arise in two- or higher-loop
approximations. However, it is not obvious whether or not
progressively including such kinds of convolution resummations
actually improves the description of the strongly nonlinear regime.
Comparisons with numerical simulations are necessary to check.
Detailed analysis of this type of extensions in the iPT is beyond the
scope of this paper, and can be considered as an interesting subject
for future work.

\subsection{Renormalized perturbation theory}

Recent progress in improving the standard perturbation theory (SPT)
was triggered by a proposition of the renormalized perturbation theory
(RPT) \cite{CS06a,CS06b}. Although this theory is formulated in
Eulerian space, there are many common features with iPT in which
resummations in terms of Lagrangian picture play an important role.
Below, we briefly discuss these common features. However, one should
note that purposes of developing RPT and iPT are not the same. The RPT
formalism mainly focuses on describing nonlinear evolutions of density
and velocity fields of matter, extrapolating the perturbation theory
in Eulerian space. The iPT formalism mainly focuses on consistently
including biasing and redshift-space distortions into the perturbation
theory from the first principle as possible. The RPT (and its
variants) is properly applicable only to unbiased matter clustering in
real space (even though there are phenomenological approaches with
freely fitting parameters, such as the model of Ref.~\cite{TNB13}, for
example). Thus, the resummation methods in RPT can be compared only
with a degraded version of iPT without biasing and redshift-space
distortions.

\subsubsection{Propagators in high-$k$ limit}

An important ingredient of RPT is an interpolation scheme between
low-$k$ and high-$k$ limits of the multi-point propagator of mass
$\varGamma^{(n)}_{\rm m}(\bm{k}_1,\ldots,\bm{k}_n)$ with $\bm{k} =
\bm{k}_{1\cdots n}$. Based on the Eulerian picture of perturbation
theory, the high-$k$ limit of the propagator is analytically evaluated
as \cite{CS06b,BCS08}
\begin{equation}
    \varGamma^{(n)}_{\rm m}(\bm{k}_1,\ldots,\bm{k}_n) \approx
    \exp\left(-\frac12 k^2 {\sigma_{\rm d}}^2 \right)
    F_n(\bm{k}_1,\ldots,\bm{k}_n),
\label{eq:3-50}
\end{equation}
in the fastest growing mode of density field, where
\begin{equation}
  {\sigma_{\rm d}}^2 = \frac{1}{6\pi^2} \int dk'\,P_{\rm L}(k'),
\label{eq:3-51}
\end{equation}
$F_n$ is the $n$th-order kernel function of SPT, and $k =
|\bm{k}_{1\cdots n}|$. Although decaying modes and the velocity sector
are also included in the original RPT formalism \cite{CS06a,CS06b}, we
neglect them for our purpose of comparison between RPT and iPT.

The multi-point propagator in the iPT has the form of
Eqs.~(\ref{eq:1-5}) and (\ref{eq:1-6}) with full orders of
perturbations. In the unbiased case, $X={\rm m}$, we have
\begin{equation}
    \varGamma^{(n)}_{\rm m}(\bm{k}_1,\ldots,\bm{k}_n) =
    \varPi(k)
    \hat{\varGamma}^{(n)}_{\rm m}(\bm{k}_1,\ldots,\bm{k}_n),
\label{eq:3-52}
\end{equation}
where $\varPi(k) = \langle e^{-i\bm{k}\cdot\bm{\varPsi}}\rangle$ is
the vertex resummation factor. The Eq.~(\ref{eq:3-52}) should also
have the same high-$k$ limit as Eq.~(\ref{eq:3-50}), since we are
dealing with the same quantities. Although explicitly proving this
property in the framework of iPT is beyond the scope of this paper, a
natural expectation arises that the high-$k$ limit of the resummation
factor $\varPi(k)$ is given by the exponential prefactor of
Eq.~(\ref{eq:3-50}), as discussed below.

In the high-$k$ limit, the factor $e^{-i\bm{k}\cdot\bm{\varPsi}}$
which is averaged over in the resummation factor strongly oscillates
as a function of displacement field $\bm{\varPsi}$. Consequently,
large values of the displacement field do not contribute to the
statistical average, and dominant contributions come from a regime
$|\bm{\varPsi}| \alt k^{-1}$. In the high-$k$ limit, this condition
corresponds to a weak field limit of the displacement field, which is
well described by the Zel'dovich approximation,
$\tilde{\bm{\varPsi}}(\bm{k}) \approx (i\bm{k}/k^2) \delta_{\rm
  L}(\bm{k})$. Assuming a Gaussian initial condition, higher-order
cumulants of displacement field in the Zel'dovich approximation are
absent in Eq.~(\ref{eq:1-6}). Since $\langle \varPsi_i \varPsi_j
\rangle_{\rm c} = \delta_{ij} \langle |\bm{\varPsi}|^2 \rangle/3$ from
rotational symmetry in real space, we have $\langle
(\bm{k}\cdot\bm{\varPsi})^2 \rangle_{\rm c} = k^2 \langle
|\bm{\varPsi}|^2 \rangle/3 = k^2 {\sigma_{\rm d}}^2$ in the Zel'dovich
approximation. Thus we naturally expect
\begin{equation}
  \varPi(k) = \left\langle e^{-i\bm{k}\cdot\bm{\varPsi}} \right\rangle
  \approx  \exp\left(-\frac12 k^2 {\sigma_{\rm d}}^2\right),
\label{eq:3-53}
\end{equation}
in the high-$k$ limit, which agrees with the exponential prefactor of
Eq.~(\ref{eq:3-50}).

Assuming that the above expectation is correct, Eqs.~(\ref{eq:3-50})
and (\ref{eq:3-52}) suggests the high-$k$ limit of normalized
propagator is given by
\begin{equation}
  \hat{\varGamma}^{(n)}_{\rm m}(\bm{k}_1,\cdots,\bm{k}_n)
  \approx F_n(\bm{k}_1,\cdots,\bm{k}_n),
\label{eq:3-54}
\end{equation}
i.e., the high-$k$ limit of the normalized propagator is given by tree
diagrams, and contributions from whole loop corrections are
subdominant. This is a nontrivial statement, since the normalized
propagator contains non-zero loop corrections in each order. For
example, taking the limit $k\rightarrow \infty$ in Eq.~(\ref{eq:2-5a})
of the one-loop approximation, we have
\begin{equation}
  \hat{\varGamma}^{(1)}_{\rm m}(k)
  \approx 1 + \frac{58}{315} \int_0^\infty
 \frac{p^2dp}{2\pi^2} P_{\rm L}(p),
\label{eq:3-55}
\end{equation}
which is apparently different from $F_1 = 1$. Actually the integral in
the RHS is logarithmically divergent for a spectrum of
cold-dark-matter type, which has an asymptote $P_{\rm L}(k) \propto
k^{-3}$ for $k\rightarrow \infty$. Thus Eq.~(\ref{eq:3-54}) does not
apparently hold when the loop corrections are truncated at any order.
Thus, Eq.~(\ref{eq:3-54}) has a highly non-perturbative nature. This
situation is natural, because the high-$k$ limit of
Eq.~(\ref{eq:3-53}) is also highly non-perturbative. When the equation
is truncated at any order, a high-$k$ limit gives divergent terms,
while the whole factor approaches to zero. The same is true for the
high-$k$ limit in the RPT formalism, Eq.~(\ref{eq:3-50}). Provided
that Eq.~(\ref{eq:3-50}) is true, Eqs.~(\ref{eq:3-53}) and
(\ref{eq:3-54}) are the same statement because of Eq.~(\ref{eq:3-52}),
which is a definition of the normalized propagator.

The above argument is readily generalized in the case of non-Gaussian
initial conditions. In the high-$k$ limit of the RPT formalism, the
exponential factor in Eq.~(\ref{eq:3-50}) is replaced by \cite{BCS10}
\begin{equation}
    \exp\left(-\frac12 k^2{\sigma_{\rm d}}^2\right) \rightarrow
    \left\langle e^{i\alpha(k)} \right\rangle
    = \exp\left(\sum_{n=2}^\infty\frac{i^n}{n!}
        \left\langle\left[\alpha(k)\right]^n\right\rangle_{\rm c}
      \right),
\label{eq:3-56}
\end{equation}
where
\begin{equation}
  \alpha(k) \equiv -i \int \frac{d^3p}{(2\pi)^3}
  \frac{\bm{k}\cdot\bm{p}}{p^2} \delta_{\rm L}(\bm{p}).
\label{eq:3-57}
\end{equation}
Comparing these equations of RPT with Eqs.~(\ref{eq:1-6}),
(\ref{eq:1-9}) of iPT, there are correspondences,
\begin{align}
  \alpha(k) &= -i \bm{k}\cdot \int \frac{d^3p}{(2\pi)^3}
  \bm{L}^{(1)}(\bm{p}) \delta_{\rm L}(\bm{p})
  = -\bm{k}\cdot\bm{\varPsi}^{(1)},
\label{eq:3-58a}\\
  \left\langle e^{i\alpha(k)} \right\rangle
  &= \left\langle
      e^{-i\bm{k}\cdot\bm{\varPsi}^{(1)}}
    \right\rangle,
\end{align}
where $\bm{\varPsi}^{(1)}$ is the linear displacement field in
configuration space at the origin. Since Eq.~(\ref{eq:3-52}) holds in
non-Gaussian initial conditions as well, the high-$k$ limit of iPT,
Eq.~(\ref{eq:3-53}), is replaced by
\begin{equation}
  \varPi(k) \approx
  \left\langle e^{-i\bm{k}\cdot\bm{\varPsi}^{(1)}} \right\rangle
  = \exp\left[
      \sum_{n=2}^\infty 
      \frac{(-i)^n}{n!}
      \left\langle
          \left(\bm{k}\cdot\bm{\varPsi}^{(1)}\right)^n
      \right\rangle_{\rm c}
  \right],
\label{eq:3-59}
\end{equation}
which agrees with the the replacement of RPT, Eq.~(\ref{eq:3-56}).
Since only the exponential factor is replaced in Eqs.~(\ref{eq:3-50})
and (\ref{eq:3-53}), the high-$k$ limit of Eq.~(\ref{eq:3-54}) does
not change even in the case of non-Gaussian initial conditions.

\subsubsection{Nonlinear interpolation I: {\sc RegPT}}

In the RPT formalism, the nonlinear propagator is approximated by
analytically interpolating the behaviors in the high-$k$ limit and the
low-$k$ limit \cite{CS06b,BCS08,CSB12}. There are at least two
prescriptions for the interpolation. An interpolation scheme of
Refs.~\cite{BCS12,TBNC12,TNB13}, which is called {\sc RegPT}, uses an
prescription for the multi-point propagator truncated at the $N$-loop
order as
\begin{multline}
  \varGamma^{(n)}_{\rm RegPT} =
  \left(
      F_n
      + \delta\varGamma^{(n)}_{\rm 1\mathchar`-loop}
      + \delta\varGamma^{(n)}_{\rm 2\mathchar`-loop}
      + \cdots
      + \delta\varGamma^{(n)}_{N{\rm \mathchar`-loop}}
      + {\rm C.T.}
  \right)
\\
  \times
  \exp\left(-\frac12 k^2 {\sigma_{\rm d}}^2\right)
\label{eq:3-60}
\end{multline}
where $\delta\varGamma^{(n)}_{M{\rm \mathchar`-loop}}$ is the $M$-loop
correction term of the propagator, and ${\rm C.T.}$ is a counterterm
to match the $N$-loop expression is exact in both limits, i.e.,
\begin{multline}
  {\rm C.T.} = 
  \left[
      \frac12 k^2 {\sigma_{\rm d}}^2  + \frac18 k^4 {\sigma_{\rm d}}^4
      + \cdots
      + \frac{1}{N!} \left(\frac{k^2 {\sigma_{\rm d}}^2}{2}\right)^N
  \right] F_n
\\
  + \left[
      \frac12 k^2 {\sigma_{\rm d}}^2 + \cdots
      + \frac{1}{(N-1)!} \left(\frac{k^2 {\sigma_{\rm d}}^2}{2}\right)^{N-1}
    \right]
  \delta\varGamma^{(n)}_{\rm 1\mathchar`-loop}
\\
  + \cdots
  + \frac12 k^2 {\sigma_{\rm d}}^2\,
    \delta\varGamma^{(n)}_{(N-1){\rm \mathchar`-loop}}.
\label{eq:3-61}
\end{multline}
The tree-level multi-point propagators are the same as the kernel
functions in SPT, i.e., $\varGamma^{(n)}_{\rm tree} = F_n$. It is
apparent that Eq.~(\ref{eq:3-60}) has the correct low-$k$ limit. In
the high-$k$ limit, we have $\delta\varGamma^{(n)}_{M{\rm
    \mathchar`-loop}} \approx (-k^2\sigma_{\rm d}^2/2)^M F_n/M!$
according to Eq.~(\ref{eq:3-50}). In this limit, it can be shown by
induction that all the loop corrections in the first parentheses of
Eq.~(\ref{eq:3-60}) including the counter term remarkably cancel each
other, leaving only the tree-level contribution $F_n$. Thus
Eq.~(\ref{eq:3-60}) also has the correct high-$k$ limit,
$\varGamma^{(n)}_{\rm RegPT} \rightarrow F_n \exp(-k^2 {\sigma_{\rm
    d}}^2/2)$ for $k \rightarrow \infty$. The {\sc RegPT} prescription
of Eq.~(\ref{eq:3-60}) can be re-expressed in a more compact form
including the counterterm as
\begin{multline}
  \varGamma^{(n)}_{\rm RegPT} =
  \left.\left[
  \exp\left(\frac12 k^2 {\sigma_{\rm d}}^2\right)\,
  \sum_{N=0}^\infty \delta\varGamma^{(n)}_{N{\rm \mathchar`-loop}}
  \right]\right|_{\rm truncated}
\\ \times
  \exp\left(-\frac12 k^2 {\sigma_{\rm d}}^2\right),
\label{eq:3-62}
\end{multline}
where $\delta\varGamma^{(n)}_{\rm 0\mathchar`-loop} \equiv F_n$, and
$[\cdots]|_{\rm truncated}$ indicates a truncation up to a given order
after completely expanding the exponential factor.

The {\sc RegPT} prescription of Eq.~(\ref{eq:3-60}) can be compared
with Eq.~(\ref{eq:3-52}) in the iPT formalism. On one hand, applying a
Taylor expansion of the resummation factor $\varPi$ and truncating at
the $n$-loop order give the same result as the $n$-loop SPT. On the
other hand, the lowest-order approximation of the resummation factor
is given by Eq.~(\ref{eq:2-3}) in real space, i.e.,
\begin{equation}
  \varPi(k) \approx \exp\left(-\frac12 k^2 {\sigma_{\rm d}}^2\right),
  \quad (k \rightarrow 0),
\label{eq:3-63}
\end{equation}
which is accidentally the same as the exponential factor in the
high-$k$ limit of Eq.~(\ref{eq:3-50}). From these observations, it is
now clear that
{\em the {\sc RegPT} prescription of Eq.~(\ref{eq:3-60}) is equivalent
  to evaluate the unbiased propagator $\varGamma^{(n)}_{\rm m}$ by
  Eq.~(\ref{eq:3-52}) in the framework of iPT, keeping only the
  lowest-order term in the vertex resummation factor $\varPi(k)$ and
  expanding all the other higher-order terms from the exponent.}
In other words, the {\sc RegPT} prescription is equivalent to the
restricted iPT formalism where the vertex resummations are truncated
at the one-loop level (without biasing and redshift-space
distortions).

\subsubsection{Nonlinear interpolation II: {\sc MPTbreeze}}

There is another scheme of interpolating the nonlinear propagators
called {\sc MPTbreeze} \cite{CSB12}, which is originally employed in
the two-point propagator in the RPT formalism \cite{CS06b}. This
method is simpler than the {\sc RegPT}, in a sense that calculations
of interpolated propagators require only one-loop integrals. In the
{\sc MPTbreeze} prescription, the interpolated propagators are given
by
\begin{equation}
  \varGamma^{(n)}_{\rm MPTbreeze}(\bm{k}_1,\cdots,\bm{k}_n)
  = F_n(\bm{k}_1,\cdots,\bm{k}_n)
  \exp\left[\delta\varGamma^{(1)}_{\rm 1\mathchar`-loop}(k)\right],
\label{eq:3-70}
\end{equation}
where the one-loop correction term of two-point propagator in the
growing mode is explicitly given by
\begin{multline}
  \delta\varGamma^{(1)}_{\rm 1\mathchar`-loop}(k)
  = \int \frac{d^3q}{(2\pi)^3}
  \frac{P_{\rm L}(q)}{504 k^3 q^5}
  \Biggl[
    6 k^7 q - 79 k^5 q^3 + 50 q^5 k^3
\\
   - 21 k q^7  + \frac34 (k^2 - q^2)^3(2k^2 + 7q^2)
    \ln\frac{|k-q|^2}{|k+q|^2}
  \Biggr].
\label{eq:3-71}
\end{multline}
The notations $P_0(k)$, $f(k)$ $\varGamma^{(n)}_\delta$ in
Ref.~\cite{CSB12} are related to our notations by $P_{\rm L}(k) =
(2\pi)^3 D_+^2(z) P_0(k)$, $\delta\varGamma^{(1)}_{\rm
  1\mathchar`-loop}(k) = D_+^2(z)f(k)$ and $\varGamma^{(n)}_{\rm
  MPTbreeze} = \varGamma^{(n)}_\delta/D_+^n(z)$, where $D_+(z)$ is the
linear growth factor. Since $\delta\varGamma^{(1)}_{\rm
  1\mathchar`-loop}(k) \rightarrow -k^2{\sigma_{\rm d}}^2/2$ in the high-$k$
limit and $\delta\varGamma^{(1)}_{\rm 1\mathchar`-loop}(k) \rightarrow
0$ in the low-$k$ limit, Eq.~(\ref{eq:3-70}) has correct limits.

It is worth noting that the prescription of Eq.~(\ref{eq:3-70})
corresponds to replacing all the loop-correction terms of propagators
by
\begin{equation}
  \delta\varGamma^{(n)}_{N{\rm \mathchar`-loop}}(\bm{k}_1,\ldots,\bm{k}_n)
  \rightarrow
  \frac{1}{N!}
  \left[
      \delta\varGamma^{(1)}_{\rm 1\mathchar`-loop}(k)
  \right]^N F_n(\bm{k}_1,\ldots,\bm{k}_n).
\label{eq:3-72}
\end{equation}
Both prescriptions of {\sc MPTbreeze} and {\sc RegPT} give similar
results, and they agree with numerical simulations fairly well in the
mildly nonlinear regime \cite{CSB12,TBNC12}. Thus the approximation of
Eq.~(\ref{eq:3-72}) turns out to be empirically good, although the
physical origin of the goodness in this prescription is somehow unclear.

According to Eq.~(\ref{eq:2-5a}) or Eq.~(\ref{eq:2-5b}), the
iPT-normalized two-point propagator of mass is related to the function
$\delta\varGamma^{(1)}_{\rm 1\mathchar`-loop}(k)$ by
\begin{equation}
  \delta\varGamma^{(1)}_{\rm 1\mathchar`-loop}(k)
  = \delta \hat{\varGamma}^{(1)}_{\rm 1\mathchar`-loop}(k)
  - \frac12 k^2 {\sigma_{\rm d}}^2,
\label{eq:3-73}
\end{equation}
where $\delta \hat{\varGamma}^{(1)}_{\rm 1\mathchar`-loop}$ is the
one-loop correction term which corresponds to the integral in
Eq.~(\ref{eq:2-5b}) without bias, $c^{(n)}_X=0$, or
\begin{equation}
  \delta\hat{\varGamma}^{(1)}_{\rm 1\mathchar`-loop}(k)
  = \frac{5}{21} R_1(k) + \frac37 R_2(k),
\label{eq:3-74}
\end{equation}
as seen in Eq.~(\ref{eq:2-104}). Substituting Eq.~(\ref{eq:3-73}) into
Eq.~(\ref{eq:3-70}), we have
\begin{equation}
  \varGamma^{(n)}_{\rm MPTbreeze} =
  F_n\,
  \exp\left[{\delta\hat{\varGamma}^{(1)}_{\rm 1\mathchar`-loop}(k)}\right]\,
  \exp\left(-\frac12 k^2 {\sigma_{\rm d}}^2\right).
\label{eq:3-75}
\end{equation}
Comparing this form with Eq.~(\ref{eq:3-62}), the relation between the
prescriptions of {\sc RegPT} and {\sc MPTbreeze} is explicit. Both
prescriptions differ in the prefactor preceding to the exponential
damping factor; a truncation scheme is employed in {\sc RegPT}, and a
simple model of the higher-loop corrections is employed in {\sc
  MPTbreeze}.

\section{\label{sec:Concl}
Conclusions
}

The iPT is a unique theory of cosmological perturbations to predict
the observable spectra of biased tracers both in real space and in
redshift space. This theory does not have phenomenological free
parameter once the bias model is fixed. In other words, all the
uncertainties regarding biasing are packed into the renormalized bias
functions $c_X^{(n)}$, and weakly nonlinear gravitational evolutions
of spatial clustering of biased tracers are described by iPT without
any ambiguity. In this way, the iPT separates the bias uncertainties
from weakly nonlinear evolutions of spatial clustering. The
renormalized bias functions are evaluated for a given model of bias.

Most of physical models of bias, such as the halo bias and peaks bias,
fall into the category of the Lagrangian bias. Redshift-space
distortions are simpler to describe in Lagrangian picture than in
Eulerian picture. The iPT is primarily based on the Lagrangian picture
of perturbations, and therefore effects of Lagrangian bias and
redshift-space distortions are naturally incorporated in the framework
of iPT.

In this paper, general expressions of the one-loop power spectra
calculated from the iPT are presented for the first time. The cross
power spectra of differently biased objects, $P_{XY}(\bm{k})$, both in
real space and in redshift space are explicitly given in terms of
two-dimensional integrals at most up to one-loop order. The final
result in real space is given by Eq.~(\ref{eq:2-14}) with
Eqs.~(\ref{eq:2-3}), (\ref{eq:2-9}), (\ref{eq:2-10}), (\ref{eq:2-11}),
and that in redshift space is given by Eq.~(\ref{eq:2-109}) with
Eqs.~(\ref{eq:2-103}), (\ref{eq:2-105}), (\ref{eq:2-106}) and
(\ref{eq:2-108}). When the vertex resummation is not preferred, one
can alternatively use Eq.~(\ref{eq:2-4}) instead of
Eq.~(\ref{eq:2-3}). An example of the renormalized bias functions is
given by Eq.~(\ref{eq:1-20}) for a simple model of halo bias.

The iPT is a nontrivial generalization of the method of
Ref.~\cite{mat08b}, which is applicable only to the case that the
Lagrangian bias is local and that the initial condition is Gaussian.
Although the derivations are quite different from each other, it is
explicitly shown that the general iPT expression of the power spectrum
exactly reduces to the expression of Ref.~\cite{mat08b} in models of
local Lagrangian bias and Gaussian initial condition.

The effects of primordial non-Gaussianity are included as well. The
consequent results are consistent with those derived by popular method
of peak-background split. In fact, the iPT provides more accurate
evaluations of the scale-dependent bias due to the primordial
non-Gaussianity \cite{mat12}. In the present paper, both effects of
gravitational nonlinearity and primordial non-Gaussianity are
simultaneously included in an expression of biased power spectrum.
Thus, the most general expressions of power spectrum with
leading-order (one-loop) nonlinearity and non-Gaussianity are newly
obtained in this paper.

In this paper, comparisons of the analytic expressions with numerical
$N$-body simulations are quite limited. In an accompanying paper
\cite{SM13}, the results in the present paper are used in calculating
the nonlinear auto- and cross-correlation functions of halos and mass,
and are compared with numerical simulations, focusing on stochastic
properties of bias. We have confirmed that effects of nonlocal bias is
small in the weakly nonlinear regime for the Gaussian initial
conditions. That is not surprising because nonlocality in the halo
bias is effective on scales of the halo mass indicated by
Eq.~(\ref{eq:1-22}); for example, $R \simeq 0.7$, $1.4$, $3.1$, $6.6\,
h^{-1}{\rm Mpc}$ for $M=10^{11}$, $10^{12}$, $10^{13}$, $10^{14}
h^{-1}M_\odot$, respectively, while one-loop perturbation theory is
applicable on scales $\gg 5$--$10\,h^{-1}{\rm Mpc}$ for $z \alt 3$.
Therefore, the predictions of iPT in Gaussian initial conditions with
one-loop approximation are almost the same as those of LRT with
Lagrangian local bias \cite{mat08a}, which have been compared in
detail \cite{SM11} with numerical simulations of halos both in real
space and in redshift space. The nonlocality of halo bias should be
important on small scales, and further investigations on the
renormalized bias functions are interesting extension of the present
work.

In the framework of iPT, the vertex resummation is naturally defined,
resulting in the resummation factor $\varPi(\bm{k})$ of
Eq.~(\ref{eq:2-3}) in real space or Eq.~(\ref{eq:2-103}) in redshift
space. The vertex resummation of iPT is closely related to other
resummation methods like RPT which are formulated in Eulerian space.
When the vertex resummation is truncated up to one-loop order, the iPT
without bias and redshift-space distortions gives the equivalent
formalism to the {\sc RegPT}, a version of RPT with regularized
multi-point propagators.

Beyond the vertex resummations, the scheme of CLPT is readily applied
to the framework of iPT as discussed in Sec.~\ref{subsec:CLPT}. Further
resummation scheme of convolution can be also considered. It might be
an interesting application of iPT to include those type of further
resummations in the presence of nonlocal bias and redshift-space
distortions.

Although the resummation technique has proven to be useful in the
one-loop approximation, it is not trivial whether the same is true in
arbitrary orders. The vertex resummation is not compulsory in iPT,
rather it is optional. The general form of vertex resummation factor
in iPT is given by Eq.~(\ref{eq:1-6}). When this exponential function
is expanded into polynomials, we obtain a perturbative expression of
power spectrum without resummation, which is an analogue to SPT.
However, for evaluations of the correlation function, the exponential
damping of the resummation factor stabilizes the numerical
integrations of Fourier transform, and therefore the vertex
resummation is preferred.

The nonlocal model of halo bias \cite{mat12} explained in
Sec.~\ref{subsec:HaloBias} is still primitive. There are plenty of
rooms to improve the model of nonlocal bias in future work. The iPT
provides a natural framework to separate tractable problems of weakly
nonlinear evolutions of biased tracers from difficult problems of
fully nonlinear phenomena of biasing.

\begin{acknowledgments}
  I thank Masanori Sato for providing numerical data of power spectra
  and correlation functions from the $N$-body simulations used in this
  paper. I acknowledge support from the Ministry of Education,
  Culture, Sports, Science, and Technology, Grant-in-Aid for
  Scientific Research (C), 21540267, 2012.
\end{acknowledgments}

\appendix


\section{\label{app:DiagRules}
Diagrammatic rules
}

\begin{figure}
\begin{center}
\includegraphics[width=18pc]{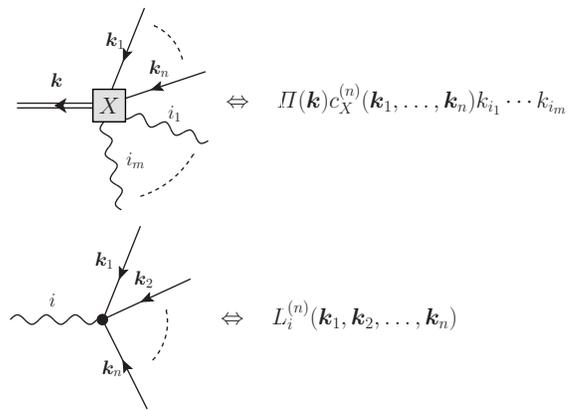}
\caption{\label{fig:iPTdiag1}
Diagrammatic rules of iPT: dynamics and biasing.
}
\end{center}
\end{figure}
\begin{figure}
\begin{center}
\includegraphics[width=15pc]{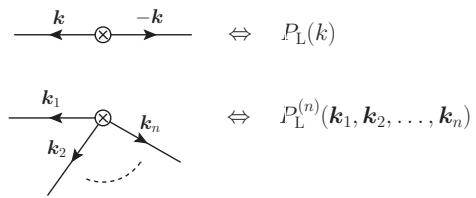}
\caption{\label{fig:iPTdiag2} 
Diagrammatic rules of iPT: primordial spectra.
}
\end{center}
\end{figure}

A set of diagrammatic rules in iPT which is used in this paper is
summarized in this Appendix. Full set of rules and their derivations
are found in Ref.~\cite{mat11}. The relevant diagrammatic rules are
shown in Fig.~\ref{fig:iPTdiag1} and \ref{fig:iPTdiag2}.
Physical
meanings of the graphs are as follows: a double solid line corresponds
to the number density field $\delta_X(\bm{k})$, a square box
represents partial resummations of dynamics and biasing, a wavy line
represents the displacement field, a black dot represents nonlinear
evolutions of the displacement field, a crossed circle represents the
primordial spectra. The procedures for obtaining a cross polyspectra
$P^{(N)}_{X_1\cdots X_N}(\bm{k}_1,\ldots,\bm{k}_N)$ of different types
of objects $X_1, \ldots, X_N$ are listed below. Auto polyspectra are
obtained by just setting $X_1 = \cdots = X_N$. The power spectrum is a
special case of polyspectra with $N=2$.

\begin{figure}
\begin{center}
\includegraphics[width=17pc]{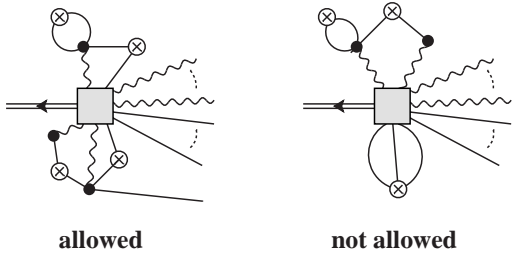}
\caption{\label{fig:VertexEx} Examples of external vertex which is
  allowed (left) and not allowed (right) in iPT. }
\end{center}
\end{figure}

\begin{enumerate}
\item Draw $N$ square boxes with labels $X_i$ ($i=1,\ldots N$), each
  of which has a double solid line. Label each double solid line with
  an outgoing wavevector that corresponds to an argument of the
  polyspectra $P^{(N)}_{X_1\cdots X_N}$.
\item Consider possible ways to connect all the square boxes by using
  wavy lines, solid lines, black dots and crossed circles, satisfying
  following constraints:
  \begin{enumerate}
    \item An end of a wavy line should be connected to a square box,
      and the other end should be connected to a black dot.
    \item An end of a solid line should be connected to a crossed
      circle, and the other end should be connected to either a square
      box or a black dot.
    \item Only one wavy line can be attached to a black dot while
      arbitrary number of solid line(s) can be attached to a black
      dot.
    \item A piece of graph which is connected to a single square box
      with only wavy lines or with only solid lines is not allowed.
  \end{enumerate}
\item Label each (solid and wavy) line with a wavevector and its
  direction. The wavevectors should be conserved at each vertex of
  square box, black dot, and crossed circle. Label each wavy line with
  spatial index together with a wavevector.
\item Apply the diagrammatic rules of Figs.~\ref{fig:iPTdiag1} and
  \ref{fig:iPTdiag2} to every distinct graphs.
\item Integrate over wavevectors as $\int d^3k_i'/(2\pi)^3$, where
  $\bm{k}_i'$ are not determined by constraints of wavevector
  conservation at vertices.
\item When there are $m$ equivalent pieces in a graph, put a
  statistical factor $1/m!$ for each set of equivalent pieces.
\item Sum up all the contributions from every distinct graphs up to
  necessary orders of perturbations.
\end{enumerate}

The rule 2.-(d) is due to partial resummations of the square box. For
example, the left diagram of Fig.~\ref{fig:VertexEx} is allowed.
There is a piece of graph that is connected to a single square box
with both wavy and solid lines. However, the right diagram of
Fig.~\ref{fig:VertexEx} is not allowed, because of double reasons. One
is that the upper piece of graph is connected to a single square box
with only wavy lines. The other is that the lower piece of graph with
only solid lines connected to a single square box. Each reason itself
prohibits this diagram from counted.


\renewcommand{\apj}{Astrophys.~J., }
\newcommand{\aap}{Astron.~Astrophys., }
\newcommand{\aj}{Astron.~J., }
\newcommand{\apjl}{Astrophys.~J.~Letters, }
\newcommand{\apjs}{Astrophys.~J.~Suppl.~Ser., }
\newcommand{\apss}{Astrophys.~Space Sci., }
\newcommand{\jcap}{J.~Cosmol.~Astropart.~Phys., }
\newcommand{\mnras}{Mon.~Not.~R.~Astron.~Soc., }
\newcommand{\mpla}{Mod.~Phys.~Lett.~A }
\newcommand{\pasj}{Publ.~Astron.~Soc.~Japan, }
\newcommand{\physrep}{Phys.~Rep., }
\newcommand{\ptp}{Progr.~Theor.~Phys., }
\newcommand{\jetp}{JETP, }


\end{document}